\documentclass[structabstract]{aa}
\usepackage{graphicx}
\usepackage{txfonts}
\usepackage{natbib}
\usepackage{color} 
\usepackage{verbatim}
\usepackage{hyperref}

\newcommand{\thickhline}{\noalign{\hrule height 0.8pt}}
 
\def\Rsolar{$R_{\odot}$}
\def\Msolar{$M_{\odot}$}

\newcommand{\peryr}{\,{\rm yr^{-1}}}
\newcommand{\Mo}{\rm{M}_{\odot}}
\newcommand{\Ro}{R_{\odot}}

\newcommand{\code}{\texttt{TRES}}

\begin{document}

\title{The evolution of stellar triples:}
\subtitle{The most common evolutionary pathways}
\titlerunning{The evolution of stellar triples}

\author{S.~Toonen \inst{1,2,3} 
\and S.~Portegies~Zwart \inst{3}
\and A.S.~Hamers \inst{4,5}
\and D.~Bandopadhyay \inst{1}} 
\institute{Institute of Gravitational Wave Astronomy, 
School of Physics and Astronomy, University of Birmingham, Birmingham, B15 2TT, United Kingdom
\\  \email{toonen@star.sr.bham.ac.uk}
\and Anton Pannekoek Institute for Astronomy, University of Amsterdam, 1090 GE Amsterdam,
The Netherlands 
\and Leiden Observatory, Leiden University, PO Box 9513, NL-2300 RA Leiden, the Netherlands 
\and Institute for Advanced Study, School of Natural Sciences, Einstein Drive, Princeton, NJ 08540, USA
\and Max-Planck Institute for Astrophysics, Karl-Schwarzschild-Str. 1, D-85741 Garching, Germany
}
\date{Received ...; Accepted ...}

\abstract
{Many stars do not live alone, but instead have one or more stellar companions. Observations show that these binaries, triples, and higher-order multiples are common. While the evolution of single stars and binaries have been studied extensively, the same is not true for the evolution of stellar triples.  }
{To fill in this gap in our general understanding of stellar lives, we aim to systematically explore the long-term evolution of triples and to map out the most common evolutionary pathways that triples go through. We quantitatively study how triples evolve, which processes are  the most relevant, and how this differs from binary evoluion.}
{We simulated the evolution of several large populations of triples with a population synthesis approach. We made use of the triple evolution code \code\, to simulate the evolution of each triple in a consistent way, including three-body dynamics (based on the secular approach), stellar evolution, and their mutual influences. We simulated the evolution of the system up until mass transfer starts, the system becomes dynamically unstable, or a Hubble time has passed. }
{We find that stellar interactions are common in triples. Compared to a binary population, we find that the fraction of systems that can undergo mass transfer is $\sim2-3$ times larger in triples. 
Moreover, while orbits typically reach circularisation before Roche-lobe overflow in binaries, this is no longer true in triples.
 In our simulations, about 40\% of systems retain an eccentric orbit. Additionally, we discuss various channels of triple evolution in detail, such as those where the secondary or the tertiary is the first star to initiate a mass transfer event. }
{}
\maketitle

\section{Introduction}

Stellar systems are the building blocks of stellar clusters and galaxies. Some of them consist of a single star such as our Solar system, but the majority of stars are members of binaries, triples, and higher order hierarchies. 
The fraction of triples is considerable. About 10\% of systems with F- \& G-type primary stars are triples.
\citep{Egg08,Rag10, Tok14b,Moe17}.
The fraction gradually increases for higher mass primaries, reaching levels of several tens of percents for early B-type and O-type stars \citep{Rem11, San14,Moe17}. The overall multiplicity fraction (that is having two or more companions)  indicates that the majority of systems with O-type primaries are triples and higher-order multiples \citep{Moe17}.

The evolution of single stars and binaries have been studied extensively, but this is not the case for the evolution of stellar triples. The evolution of triples iss complicated as it is a combination of stellar evolution, stellar interactions, and three-body dynamics \citep[see][for a review]{Too16}.
An example of the latter are Lidov-Kozai (LK) cycles \citep{Lid62,Koz62} in which secular gravitational effects induce eccentricity oscillations in the inner binary system \citep[see][for a comprehensive review]{Nao16}. 
However, realistic stellar triples are not purely gravitational systems  because stars are objects with finite sizes and lifetimes and they have non-static natures (e.g. evolving masses and radii). Due to stellar evolution, a triple can move in or out of the regime where three-body dynamics modifies the system \citep[e.g.][]{Fab07, Per12, Sha13, Ham13, Mic14,Ant17, Too18b, Ham19b, Ham19c, Fra19, Fra19b}. 
Therefore, to study the evolution of stellar triples consistently, stellar evolution and a full treatment of dynamics need to be taken into account simultaneously. Here we present the systematic exploration of the evolution of stellar triples with low and intermediate mass stars. 
We address what the common evolutionary channels are that these triple systems evolve through and estimate how common these channels are. In particular we discuss which evolutionary pathways are open to triples that are not open to isolated binaries. We address these questions using a population synthesis approach with the triple evolution code \texttt{TRES} \citep{Too16}. In Sect.\,\ref{sec:met} we outline our model assumptions for the initial populations of triples and their evolution. We discuss the five most common evolutionary channels and their demographics in Sect.\,\ref{sec:res}, as well as their dependence on our model assumptions in  Sect.\,\ref{sec:discussion}. We summarise our findings in Sect.\,\ref{sec:concl}.

\section{Method}
 \label{sec:met}

\begin{table*}[ht!]
\centering
\caption{Distributions of the initial binary parameter mass, mass ratio, orbital separation and eccentricity.}
\begin{tabular}{lccc}
\thickhline
Parameter & Model OBin & Model T14  & Model E09\\
\thickhline
Mass of binary primaries & Kroupa IMF $^{(1)}$& Kroupa IMF $^{(1)}$& Kroupa IMF $^{(1)}$  \\
Mass ratio & Uniform  & Uniform & Model E09$^{(2)}$\\
Orbital separation & Uniform in log($a$)$^{(3)}$ & Log-normal$^{(4,5,6)}$ & Model E09$^{(2)}$\\
Eccentricity & Thermal distribution $^{(7)} $ & Thermal distribution $^{(7)} $& Thermal distribution $^{(7)} $\\
Inclination & Uniform in cos($i$) & Uniform in cos($i$) & Uniform in cos($i$)  \\
Argument of pericentre & Uniform& Uniform& Uniform\\
\hline
\end{tabular}
\label{tbl:init_param}
\begin{flushleft}
\tablefoot{$^{(1)}$ \citet{Kro93}; 
$^{(2)}$\citet{Egg09};
$^{(3)}$\citet{Abt83}; 
$^{(4)}$\citet{Duq91}; 
$^{(5)}$\citet{Rag10};
$^{(6)}$\citet{Tok14b};
$^{(7)}$ \citet{Heg75}
}
\end{flushleft}
\end{table*} 
 
To construct a map of the most common evolutionary pathways through which triples evolve, we simulated the evolution of representative populations of triple star-systems. We focus on isolated, hierarchical, coeval, initially dynamically stable systems. The simulation of an individual triple is ceased if a Hubble time has passed (here considered to be 13.5Gyr),  if the system becomes dynamically unstable, or if mass transfer commences.
We describe the primordial triple populations adopted in this study in Sect.\,\ref{sec:init_distr_tr} , and  our computational method using the triple population synthesis code the code \code in Sect.\,\ref{sec:code}.

\subsection{Primordial population}
\label{sec:init_distr_tr}
When a triple is sufficiently hierarchical, it can be described by an inner binary whose centre of mass is orbited by a distant star, that is the outer star. The masses of the stars in the inner orbit are $m_1$ (primary mass) and $m_2$ (secondary mass), where $m_1 \geqslant m_2$ initially. The mass of the outer star is $m_3$ (tertiary mass). We define the inner and outer mass ratios as $q_{\rm in } \equiv m_2/m_1$ and $q_{\rm out} \equiv m_3/(m_1 +m_2)$.
We note that the evolution of a star is primarily determined by its mass, with secondary effects due to metallicity (here taken as solar) and rotation\footnote{Stellar rotation is known to affect the evolution of massive stars \citep[e.g.][]{Mae00}, and in extreme cases it may lead to chemically homogeneous evolution. For the mass range studied in this paper, rotation is expected to be less important and therefore not taken into account here.}.
The inner and outer orbits are given by the semimajor axis ($a_{\rm in}$ and $a_{\rm out}$), the eccentricities ($e_{\rm in}$ and $e_{\rm out}$), and the arguments of pericentre ($g_{\rm in}$ and $g_{\rm out}$). The relative orientation of the two orbits with respect to one another is measured by the mutual inclination $i$. The longitudes of ascending nodes describe the orientation on the sky, but do not affect the evolution of the system \citep[e.g.][]{Har68, Nao13}. The stellar rotation is described through the angular velocity $\Omega_1,\Omega_2,\Omega_3 $ for the primary, secondary and tertiary, respectively.

We initialise our simulations with a population of stellar triples on the zero-age main-sequence (ZAMS). 
We apply three different models, named OBin, T14 and E09.  
The model OBin is based on our understanding of observed populations of primordial binaries, whereas model T14 and  model E09 are based on triples as proposed by 
\citet{Tok14b} and \citet{Egg09}. The sample of \citet{Tok14b} consists of 4847 FG-dwarfs 
within 67 pc, and the one used in \citet{Egg09} is a magnitude-limited sample of 4558 stellar systems. 
The models are described in detail below and an overview is given in Tbl.\,\ref{tbl:init_param}. 

For all models, primary masses are drawn from an initial mass function  based on \citet{Kro93} in the range of 1-7.5$\Mo$. In this mass range single stars evolve to the remnant stage within a Hubble time and do not experience a supernova explosion. In all models and their normalisation we require the primary masses to be above $0.08\Mo$ and companion masses above $0.008\Mo$ to exclude sub-stellar components. To calculate event rates, we assume that 15\% of stellar systems are triples, and 50\% are binaries \citep{Tok14b}. Furthermore, for Galactic rates we assume a constant star formation history of 3 $\Mo$ per year for 10 Gyrs.

  \begin{figure}[ht!]
    \centering
	\includegraphics[width=\columnwidth]{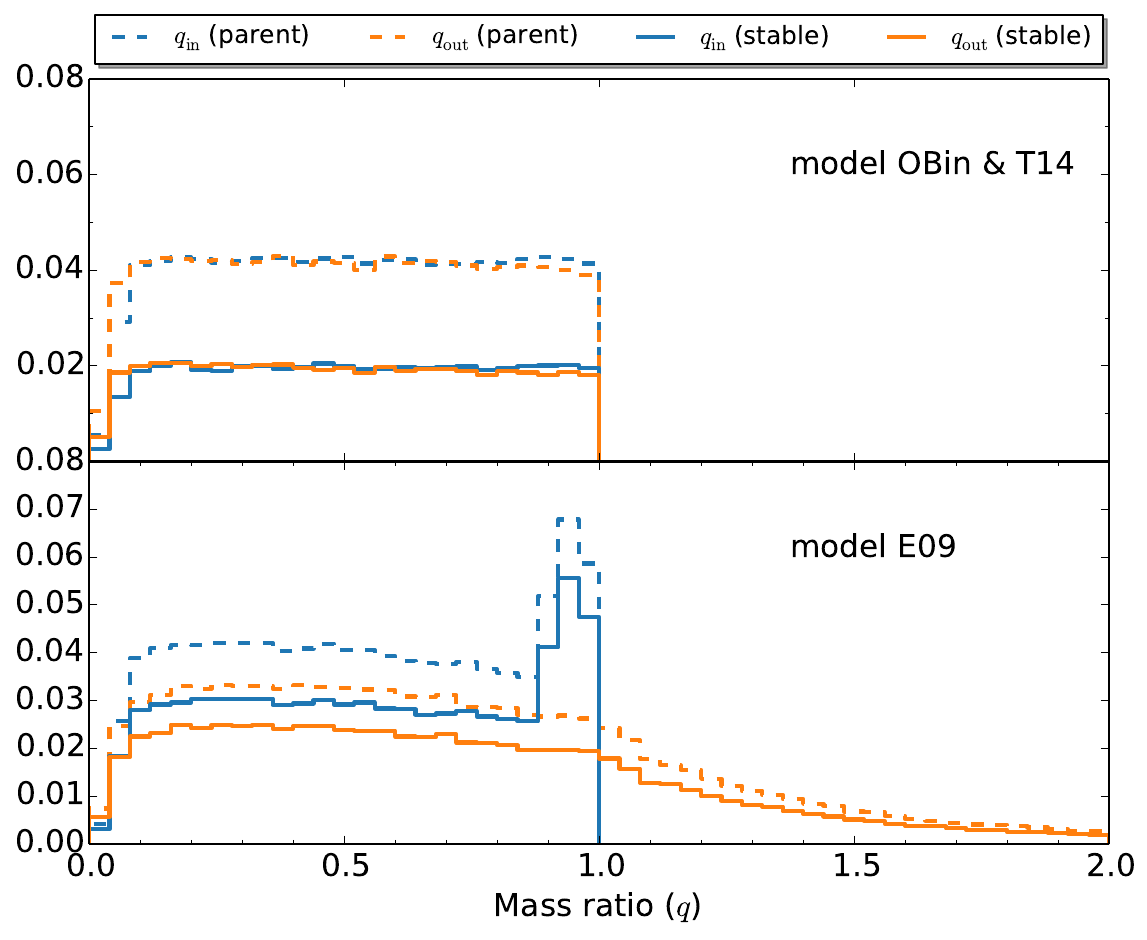} 
    \caption{Binned distributions of initial mass ratios for the inner and outer orbit (blue and orange lines). The dashed lines (normalised to unity) reflect the parent distributions from which we draw the mass ratios. The solid lines show the subset of systems that are dynamically stable according to the criterion by \cite{Mar99} and that are used in the simulations. On the top, a flat mass ratio distribution is shown as used in model OBin and T14, whereas on the bottom the mass ratio distribution from model E09 is shown.  }
    \label{fig:init_q}
    \end{figure}

The mass ratios of model OBin and Tok14 (upper panel of Fig.\,\ref{fig:init_q} ) are based on our understanding of the mass ratio distribution of ZAMS binaries. These binaries exhibit mass ratios that are unbiased with respect to the primary and secondary mass, that is a flat mass ratio distribution between 0 and 1 \citep[e.g.][]{Rag10, San12,Duc13,  Moe17}. 
We note that the mass ratio distributions decrease sharply at small mass ratios, as these systems would contain substellar companions.  
For model E09, the mass ratio distribution of the inner binaries (bottom panel of Fig.\,\ref{fig:init_q}) is roughly flat except for an enhancement at nearly equal-mass stars. 
The mass ratio distribution of the outer binaries is also roughly flat for $0< q \leqslant 1$, however, there is a tail that extends to high mass ratios. In model E09 the tertiary mass is not limited to the total mass of the inner binary. As a consequence, model E09 predicts $\sim10$ times more systems with $m_3>7.5\Mo$  compared to model OBin \& Tok14 (which are not included in this study, see also Tbl.\,\ref{tbl:birthrates}).

\begin{table*}
\caption{
Birthrates of stellar triples per solar mass of stars (middle column) and in the Milky Way (right column). The bottom panel shows the occurrence rates of the most common evolutionary pathways of intermediate mass triples (that is initially $1< m_1\leq 7.5\Mo$ and $m_2,m_3<7.5\Mo$) based on a population synthesis simulation including a realistic mix of single, binary and triple stars (see Sect.\,\ref{sec:met}). }
\begin{tabular}{ll|ccc|ccc}
&& \multicolumn{3}{c|}{Total rate (per $10^4\Mo$ )} & \multicolumn{3}{|c}{Galactic rate ($10^{-3}\peryr$)}\\
&& OBin & T14  & E09 & OBin & T14  & E09\\
\hline 
Total birthrate  && 1740 & 1725 & 1489	 &  522 &  517  & 45\\
 \multicolumn{2}{l|}{- with low-mass ($m_1\leq1\Mo$)} & 1563& 1544& 1294	& 469 &463 & 388\\
  \multicolumn{2}{l|}{- with intermediate-mass ($1< m_1\leq 7.5\Mo$)} & 167& 168& 168	& 50 &51 & 51\\
  \multicolumn{2}{l|}{- with intermediate-mass ($1< m_1\leq 7.5\Mo$, $m_2,m_3<7.5\Mo$)} & 165& 166& 154	& 50 & 50 & 46\\
  \multicolumn{2}{l|}{- with high-mass ($m_1>7.5\Mo$)} & 13& 13& 27 	& 3.8 & 3.8 & 8.2 \\
\hline 
\multicolumn{2}{l|}{Birthrate per evolutionary channel} & 		&         & 	&&&\\
Mass transfer initiated by: &
primary star &     				  105       &         125 	&   106        &   30  &  35   & 30             \\
& secondary star & 				1.3      &           1.2 	&   1.1       &   0.3 & 0.3 & 0.3          \\
& tertiary star  & 				0.9   &             1.9 	&   1.4       &   0.3 & 0.6 & 0.4          \\
Dynamical instability &&  		6.9      &           3.8 	&   3.8       &   2.0 &  1.1  & 1.1               \\
No interaction &&  				40      &           34 	&   42       &   17  &   13   & 15             \\
\hline 
\end{tabular}
\label{tbl:birthrates}
\end{table*}

 \begin{figure}[h!]
    \centering
	\includegraphics[width=\columnwidth]{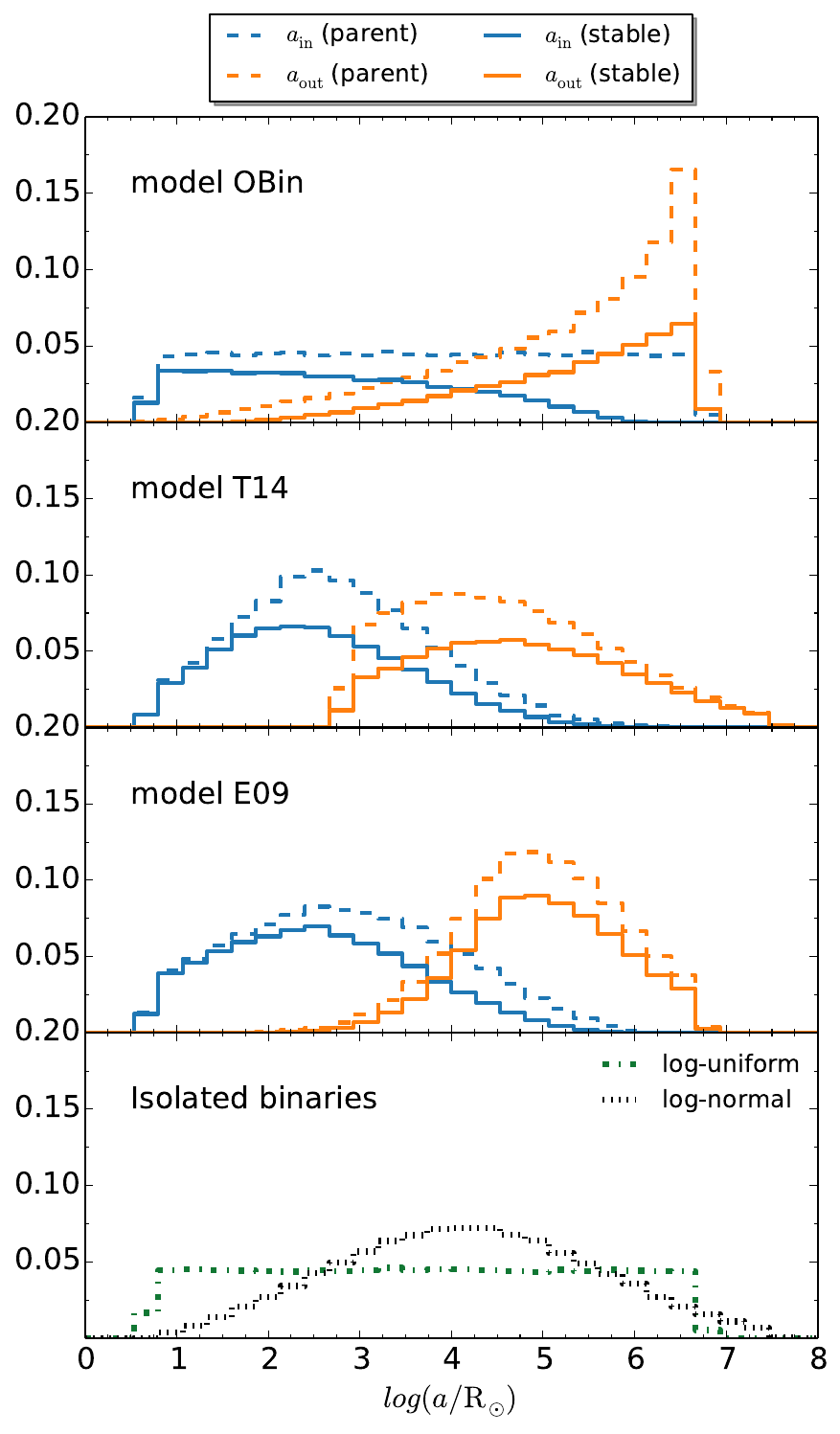} 
    \caption{ Binned distributions of the initial inner and outer semimajor axis (blue and orange lines).  The dashed lines (normalised to unity) reflect the parent distributions from which we draw the semimajor axes. The solid lines shows the subset of systems that are dynamically stable and that are used in the simulations. From top to bottom model OBin, T14 and E09 are shown. The bottom panel shows the distribution of semimajor axes that are commonly used for isolated binaries.  }
    \label{fig:init_a}
    \end{figure} 
 
Next we describe the distributions of the inner and outer orbital separations (Fig.\,\ref{fig:init_a}). For model OBin, we adopt a flat distribution in the logarithm of the orbital separation \citep[e.g.][]{Abt83, Kou07}. The distribution of orbital separations ranges from 5\Rsolar\,to $5\cdot 10^6$\Rsolar.
Even though the outer orbital separation is drawn from the same distribution as for the inner orbit, the distribution is biased to large orbital separations as we require $a_{\rm out}>a_{\rm in}$. 
The flat distribution is a classical assumption in binary population synthesis studies \citep[e.g.][]{Yun93,deK93, Por98, Nel01, Hur02, Bel02, Han02,  Izz09, Rui09,Men10, Too12,  Cla14, Haa13, Too18}. It is a first-order approximation of the observed distribution of binary separations. However modifications have been proposed in which the shape of the distribution changes as a function of primary star mass \citep[e.g.][]{Duc13, Moe17}. For example for Solar-type stars a log-normal distribution is preferred \citep[e.g.][]{Duq91, Rag10}, while a power-law with moderate slope is preferred for O- and B-type stars \citep{San12, San14}.

In the second panel from the top in Fig.\,\ref{fig:init_a}, we show the orbital separations for model T14. In this model a log-normal distribution of periods ($P$ in days) is assumed with mean $\mu = 5$ and dispersion $\sigma=2.3$  \citep{Duq91, Rag10, Tok14b}. 
Lastly, the primordial semimajor-axes of model E09 are shown in the third panel from the top in Fig.\,\ref{fig:init_a}. The inner orbital separations are similar to those of the log-normal distribution in model T14, but the outer orbits are generally wider by a factor of a few.

After drawing the inner and outer orbital separation according to the distributions described above (hereafter parent distributions), we require that the triple is dynamically stable. We use the stability criterion as given by \cite{Mar99}, see also Eq.\,\ref{eq:stab_crit}. The resulting distributions of stable systems are shown as solid lines in Fig.\,\ref{fig:init_q}~and~Fig.\,\ref{fig:init_a}. Comparing the populations before and after the stability cut shows that there are no signs of a strong bias with inner or outer mass ratio, however, the resulting distributions are biased to larger ratios of outer to inner semimajor axis. As a result, the inner orbit is biased towards shorter periods, and the outer orbit towards longer periods. We neglect the (dynamical and stellar) evolution on the pre-MS, and assume that on the zero-age MS the initial conditions are as described above \citep[but see e.g.][]{Moe18}.

  \begin{figure}
    \centering
	\includegraphics[width=\columnwidth]{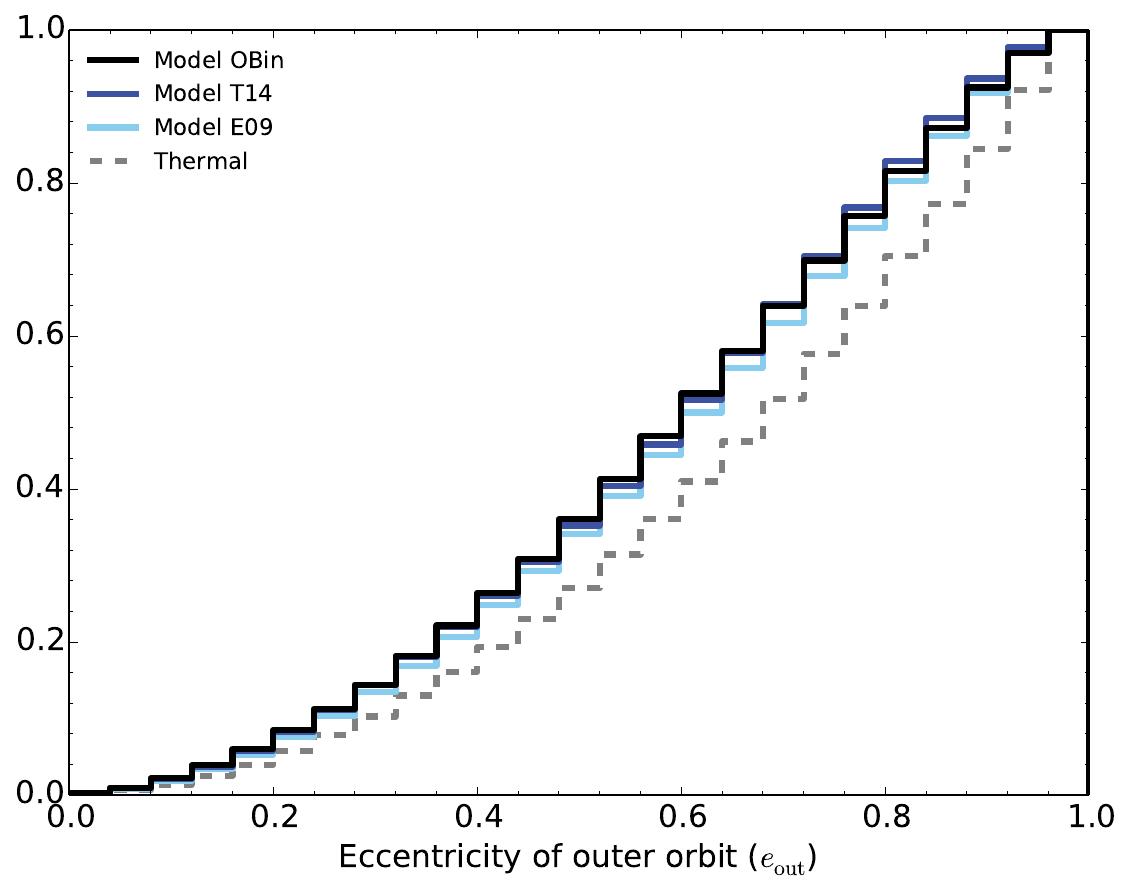} 
    \caption{Cumulative distribution of the initial eccentricity of the outer orbit normalised to unity. The dashed line shows the parent distribution \citep[i.e. thermal distribution,][]{Heg75}, and the solid lines represent the dynamically stable systems from each model used in our simulations. }
    \label{fig:init_eout}
    \end{figure}

For all models, we assume a thermal distribution of eccentricities \citep{Heg75} between 0 and 1. 
 After removing the dynamically unstable systems the distribution of outer eccentricities is significantly affected (Fig.\,\ref{fig:init_eout}). The fraction of highly eccentric outer orbits is reduced, and overall the distribution has flattened. This is in line with the observed eccentricity distribution of wide binaries with projected separations larger than 50 au \citep{Tok16} based on the same 67pc sample as model T14. Generally observations show that the eccentricity distributions are less steep than the thermal distribution \citep{Rag10, Tok16, Moe17}.

For the arguments of pericentre we adopt a uniform distribution between $-\pi$ and $\pi$, and for the inclination a uniform distribution in the cosine of the inclination between 0 and $\pi$.  \citet{Tok17} finds that this is a good approximation for outer projected separations $\gtrsim 10^5\Ro$, but at $\lesssim 10^4\Ro$ there is more alignment (see also Fig.\,\ref{fig:init_a}). The orbit alignment is stronger for systems with low-mass primaries ($m_1<1.2\Mo$, where the median initial primary mass in our simulation is 1.65\Msolar.)
For the angular velocities of the initial stellar rotation, we follow \cite{Hur00}.

\subsection{Modelling of triple evolution with \code}
\label{sec:code}

In \citet{Too16}, we presented a code to simulate the evolution of hierarchical triple star systems; stellar evolution is combined with three-body dynamics. Single stellar evolution is included through the fast stellar evolution code \texttt{SeBa} \citep{Por96, Too12}. 
 The code uses analytical formulae based on detailed single star tracks from \cite{Hur00} and provides parameters such that stellar radius and wind mass loss rates as a function of time.
The orbits are solved using the secular approach \citep{Har68, For00, Nao13}. The method is based on \cite{Ham13} and solves a set of first-order ordinary differential equations including secular three-body dynamics (accurate up to and including the octupole order), tides, general relativistic effects \citep{Pet64}, mass loss and mass transfer. Furthermore, precession from general relativistic effect \citep{Bla02}, due to tides acting on the stars \citep{Sme01}, and due to stellar rotation \citep{Fab07} are included. General tidal damping is included for stars with convective envelopes through equilibrium tides, with radiative envelopes through dynamical tides, and degenerate stars separately following \citet{Hur02}.
We adopt the Roche lobe equivalent for an eccentric orbit as derived by \cite{Sep07} which is a function of eccentricity as well as the rotation of the stars. 
We assume that stellar winds are emitted spherically symmetrically and fast compared to the orbit, such that the effect on the orbit is adiabatic, and the orbit widens as $\dot{a}/a = -\dot{M}_{\rm total}/M_{\rm total}$ \citep[e.g.][]{Hua56, Hua63}. Furthermore, we assume the wind matter is not accreted by the stellar companions.

\code\ is written within the Astrophysics Multipurpose
Software Environment, or AMUSE \citep{Por09, Por13, Por18}.  This is a component library
with a homogeneous interface structure and can be downloaded
for free at \href{http://amusecode.org/}{{\color{blue}amusecode.org}}.
In the AMUSE framework new and existing code from different domains can be easily coupled (e.g. stellar dynamics, stellar evolution, hydrodynamics and radiative transfer). 
As a result, the triple code can be easily extended in the future to include a detailed stellar evolution code or a direct N-body code to solve the dynamics of unstable triples.

 \begin{figure*}
    \centering

    \begin{tabular}{cc}
	\includegraphics[width=\columnwidth]{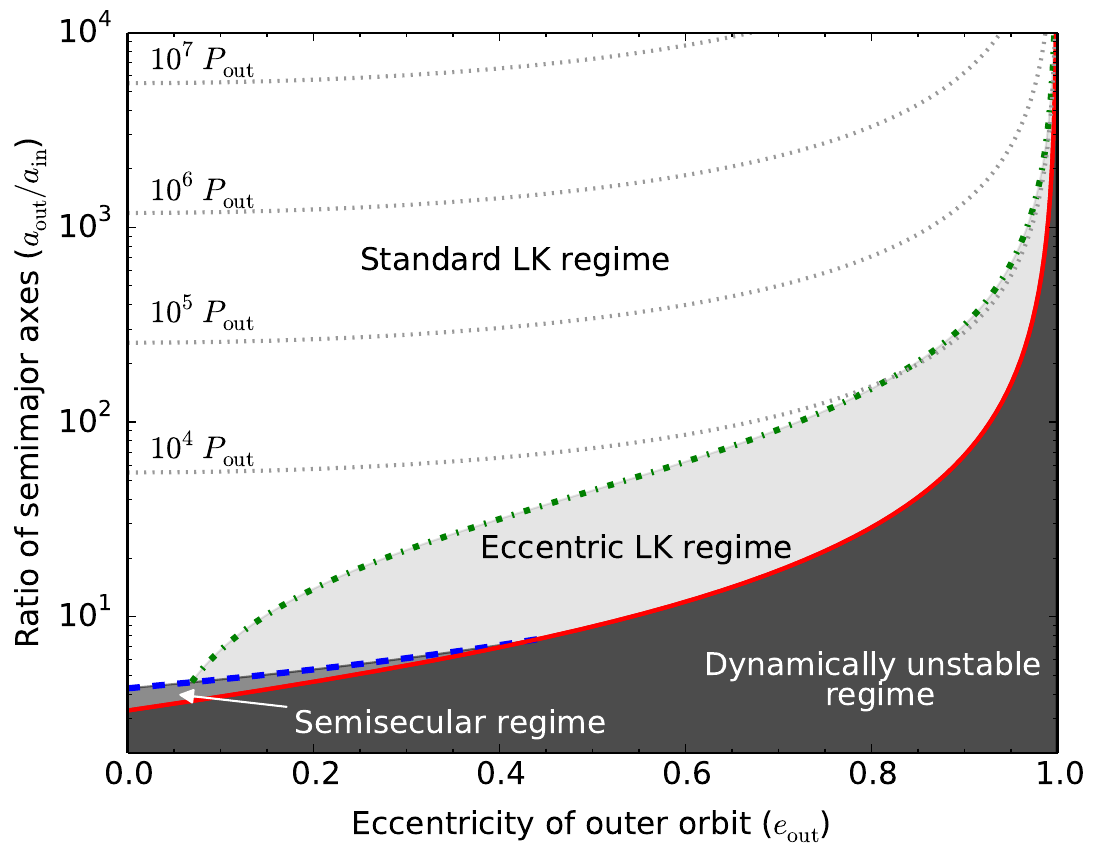}  & 
	\includegraphics[width=\columnwidth]{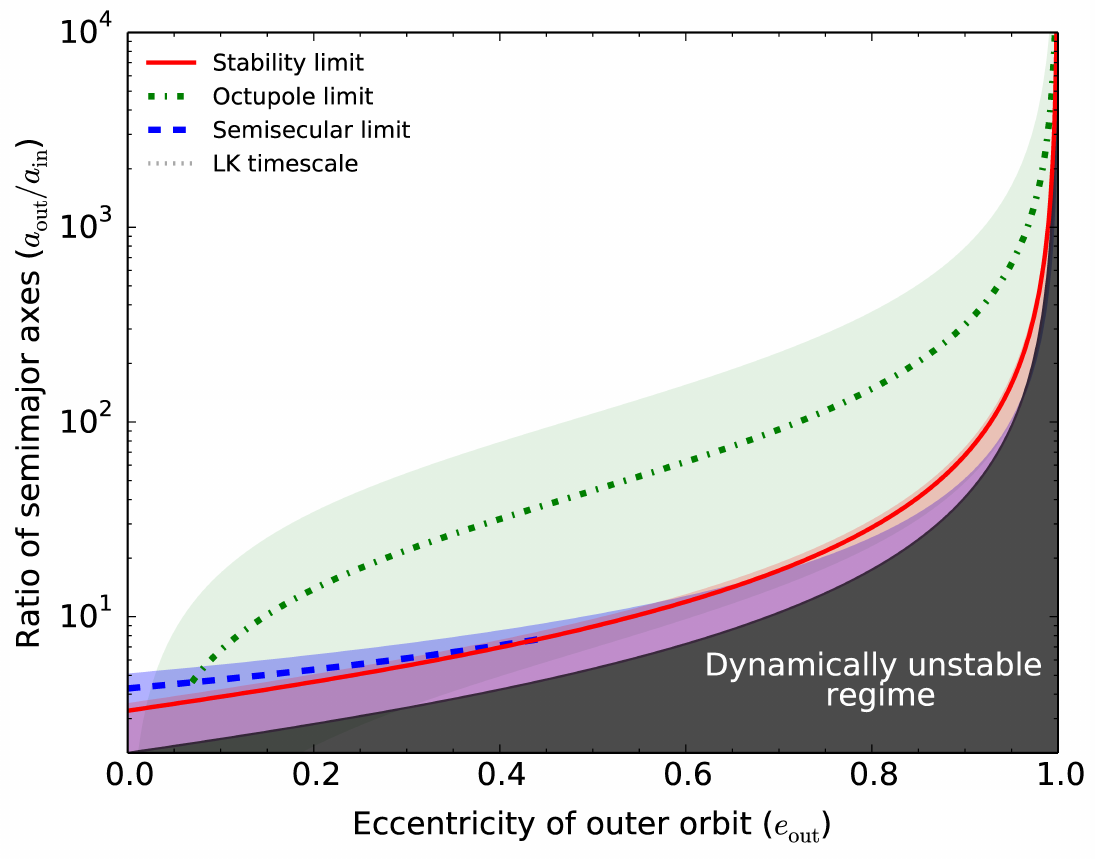} 
	\end{tabular}
    \caption{Visualisation of the parameter space of the different dynamical regimes for triple systems (Sect.\,\ref{sec:code}).  On the left a triple with  $m_1=1\Mo$, $m_2=0.5\Mo$, $m_3=0.75\Mo$, and $i=0^\circ$ is assumed. The black dotted lines represent constant Lidov-Kozai timescales of value $10^4P_{\rm out},10^5P_{\rm out},10^6P_{\rm out},$ and $10^7P_{\rm out}$. The higher-order (octupole) term of the three-body interaction is important below the green dash-dotted line assuming a limit at $\epsilon_{\rm oct} = 0.005$. The triple is in the semisecular regime below the blue dashed line - assuming the inner eccentricity is 0.99 or can be reached during the LK cycles. For lower values of $e_{\rm in}$ the line moves downwards. The system is not dynamically stable below the red solid line. Below the blue dashed and red solid line the secular approach breaks down. On the right, the shaded areas represent how the boundary of the corresponding colour moves for different stellar masses, assuming $m_2, m_3$ in $[0.1,0.9]\Mo$. For the red shaded area we also take into account how the boundary of dynamical stability shifts when the inclination varies between $0-180^\circ$.  }      
    \label{fig:theory}                                      
    \end{figure*}

For a full description of the \code\,and an overview of relevant processes in triple evolution, see \cite{Too16}. In the following we highlight a number of processes.

\subsubsection{Dynamical stability} 
In this paper we focus on dynamically stable systems. Dynamically unstable systems tend to be short-lived and typically lead to dissociation or ionisation of the systems \citep[e.g.][]{Van07}.  
Here we apply the stability criterion of  \citet{Mar99, Mar01}:
\begin{eqnarray}
\frac{a_{\rm out}}{a_{\rm in}}|_{\rm crit} = \frac{2.8}{1-e_{\rm out}} \left( 1-\frac{0.3i}{\pi} \right) \cdot 
\left( \frac{(1+q_{\rm out})\cdot(1+e_{\rm out})}{\sqrt{1-e_{\rm out}}} \right)^{2/5}. 
\label{eq:stab_crit}
\end{eqnarray}
 Triples are unstable if $\frac{a_{\rm out}}{a_{\rm in}} < \frac{a_{\rm out}}{a_{\rm in}}|_{\rm crit}$. 
 
\subsubsection{Lidov-Kozai cycles} The lowest-order manifestation of three-body dynamics are the LK cycles \citep[i.e. quadrupole order,][]{Lid62, Koz62, Nao16}. Oscillations are induced in inclination and eccentricity of the inner binary system with a timescale of  approximately \citep{Kin99, Ant15}:
  \begin{equation}
t_{\rm kozai} =  \alpha  
\frac{P_{\rm out}^2}{P_{\rm in}} 
\frac{m_1+m_2+m_3}{m_3} \left(1-e_{\rm out}^2\right)^{3/2},
\label{eq:t_kozai}
\end{equation}
where $P_{\rm in}$ and $P_{\rm out}$ are the periods of the inner and outer orbit, respectively. $\alpha$ is a dimensionless quantity of order unity that depends weakly on the mutual inclination, inner eccentricity and argument of periastron. 
Within the test-particle approximation and circular initial orbits, the maximum eccentricity $e_{\rm max}$ is a function of the initial mutual inclination  $i_{\rm i}$ as \citep{Inn97}:
\begin{equation}
e_{\rm max} = \sqrt{1-\frac{5}{3} \mathrm{cos}^2(i_{\rm i})},
\label{eq:e_max}
\end{equation}
In this approximation the LK cycles only take place when the initial inclination is between 39.2-140.8$^{\circ}$. 
When dropping some of the simplifications of this approximation (for example with non-zero initial eccentricities, realistic stellar masses ratios), the range of initial inclinations expands, higher-order levels of three-body dynamics can become important, and richer dynamical behaviour is expected. 
\subsubsection{Eccentric Lidov-Kozai regime} 
The next order of approximation in the secular approach, the so-called octupole level, gives rise to more extreme dynamical interaction. In the octupole level, the system can experience a flip in its
inclination (that is transition from prograde to retrograde or vice versa) or modulations in the eccentricity of the inner orbit that can reach close to unity \citep[e.g.][]{For00, Bla02,Lit11, Nao13, Sha13, Tey13}. 
See also \cite{Nao16} for an extensive review of the eccentric LK regime.
Generally the octupole term is of importance if the octupole parameter $|\epsilon_{\rm oct}| \gtrsim 0.001-0.01$ \citep{Lit11, Kat11, Tey13,Li14}, where  
\begin{equation}
\epsilon_{\rm oct} = \frac{m_1-m_2}{m_1+m_2} \frac{a_{\rm in}}{a_{\rm out}} \frac{e_{\rm out}}{1-e_{\rm out}^2}.
\label{eq:e_oct}
\end{equation}

\subsubsection{Semisecular evolution} 
In the semisecular regime \citep[e.g.][]{Ant12, Kat12, Luo16} the triple is technically dynamically stable, but the hierarchy is moderate;  the perturbations from the outer star on the inner orbit occur on a timescale that is comparable to or shorter than the dynamical times of the inner or outer binary. The regime is linked to rapid oscillations in the inner eccentricity and stellar collisions \citep[e.g.][]{ Kat12, Set13,Ant14, Ant14b, Bod14, Hai18}. 
The orbit-average treatment breaks down when: 
\begin{equation}
\sqrt{1-e_{\rm in}} \lesssim 5\pi q_{\rm out} \left ( \frac{a_{\rm in}}{a_{\rm out}(1-e_{\rm out})} \right)^3,
\label{eq:orbit_average}
\end{equation}
as derived by \citet{Ant14} assuming the instantaneous quadrupole
order torque, maximal torque at the periapsis approach and $e_{\rm in} \rightarrow 1$.
 The current study is based on the secular approach and does not consider the semisecular regime\footnote{But see for example \cite{Too17} for a study of wide triples that evolve into the semisecular regime due to stellar evolution, and lead to stellar collisions and transients on timescales of Gyrs.}. For the subset of moderate hierarchical systems, the maximum inner eccentricity is even higher than expected from secular theory, and likely leads to an earlier start of RLOF.

\subsection{Visualisation of the dynamical regimes }
The four processes described above give rise to different dynamical regimes in which each of them dominate over the other processes. Fig.\,\ref{fig:theory} aims to discerns the parameter space of the dynamical regimes; for a single triple in Fig.\,\ref{fig:theory}a, whereas Fig.\,\ref{fig:theory}b aims to show how the boundaries between the regimes vary for triples with different mass ratios. In general, systems with $a_{\rm out}(1-e_{\rm out})/a_{\rm in}$ smaller than a factor of a few are dynamically unstable. If the ratio is slightly larger the system is in the eccentric LK regime. If however, the inner eccentricity is (can become) large, and consequently the outer eccentricity is small, the system is in the semi-secular regime. For even larger ratios, the systems experience standard LK cycles, with ever-larger periods up to and beyond a Hubble time.

\begin{figure}
	\includegraphics[width=\columnwidth]{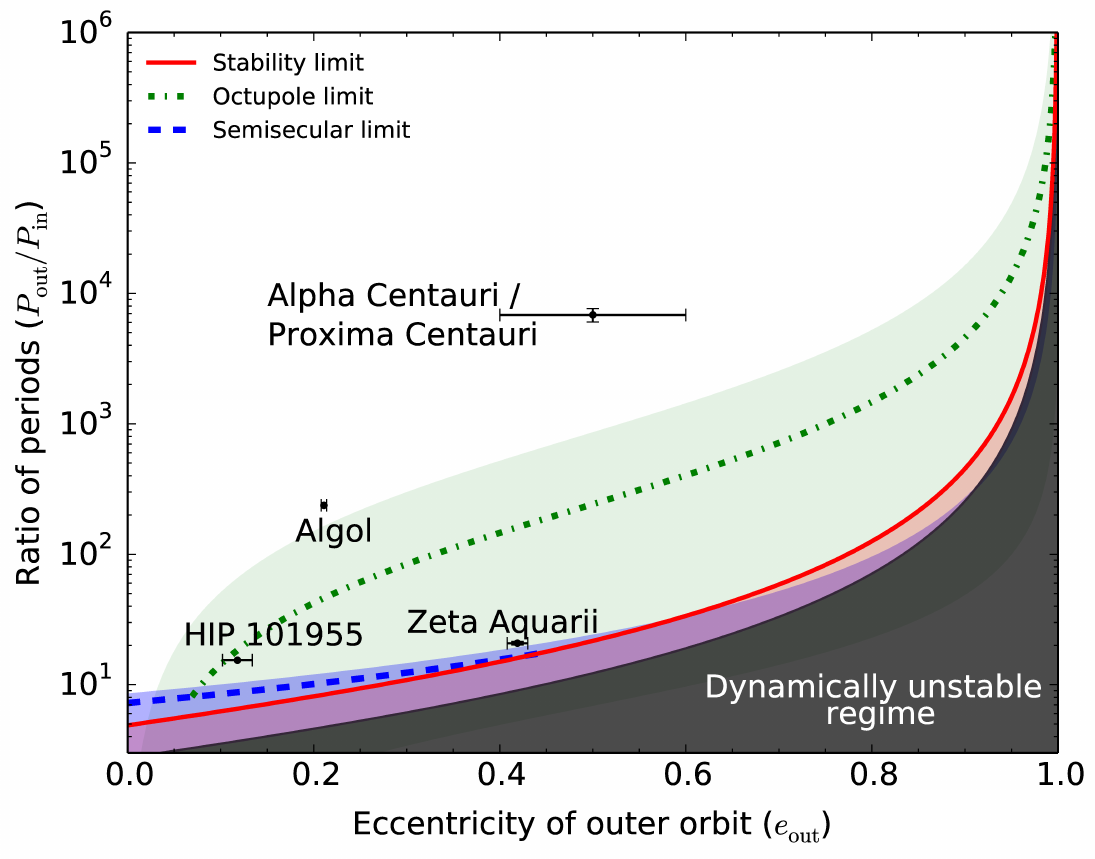}    
    \caption{Visualisation of the parameter space of the different dynamical regimes with several known triples.  The figure is similar to Fig.\,\ref{fig:theory}b but with $P_{\rm out}/P_{\rm in}$ on the y-axis in stead of $a_{\rm out}/a_{\rm in}$ because these are better measured in the current sample of overplotted observed triples overplotted. Depending on the method with which the orbit is studied, either the period or the semi-major axis may be determined more accurately. }      
    \label{fig:theory_p}                                      
    \end{figure}

To give a few examples, we have overplotted a number of known triples in Fig.\,\ref{fig:theory_p}. 
The nearest neighbour to the Sun is the stellar triple consisting of the Alpha Centauri system and Proxima Centauri. 
Within the uncertainties in $e_{\rm out}= 0.50^{0.08}_{-0.09}$ 
\citep{Ker17}, the system is far away from the eccentric LK regime, the semisecular regime, and the dynamically unstable regime. Even if the inclination is favourable for LK cycles, the timescale is orders of magnitude larger than the outer period of approx 0.55Gyr.

Algol is a post-mass transfer system with an inclination around 95$^{\circ}$ \citep{Kis98,Bor10}. The system is in the standard LK regime, but close to the eccentric LK-regime. The boundary of the eccentric LK-regime depends on the masses of the triple, as indicated by the green shaded area (Eq.\,\ref{eq:e_oct}). Specifically $a_{\rm out}/a_{\rm in}|_{\rm crit} \propto \mathcal{Q}\equiv (m_1-m_2)/(m_1+m_2)$. 
The boundary of the eccentric LK-regime for Algol lies between the green dash-dotted line and the upper boundary of the green area, with $\mathcal{Q}\sim0.64$, $\mathcal{Q}=\frac{1}{3}$, and $\mathcal{Q}=0.8$ respectively \citep{Zav10, Zav17}. And so, the octupole term is likely not a dominant factor in the evolution of Algol. As the inclination is high, LK-cycles are expected \citep[see also][]{Kis98,Bor04,Bor10}.

HIP101955 is a triple with well-determined orbital parameters \citep[see e.g.][]{Tok17b} in the green shaded area in Fig.\,\ref{fig:theory_p}. However, given its masses $m_1=0.74\Mo$,$m_2=0.62\Mo$, $\mathcal{Q}\approx 0.09$ \citep{Tok17b}, the boundary of the eccentric LK-regime  lies below the green dashed line ($\mathcal{Q}=\frac{1}{3}$) by a factor $\frac{1}{3}/0.009 \approx 3.7$, such that HIP101955 is close to the eccentric LK regime, but likely not strongly affected by the octupole term. The same holds for the semi-secular regime. Strong LK cycles are expected with short timescales ($\approx 560$ years) that might be observable \citep{Mal07, Xu15}. 

 Finally Zeta Aquarii is an interesting system as it lies close to the limit of dynamical stability. Given its high inner mass ratio ($m_1, m_2 \approx 1.4\Mo$) the octupole parameter is small \citep{Tok16b}, such that the octupole term is likely not very important. Regarding the semi-secular regime, as the outer mass ratio is small $q_{\rm out}\approx 0.6/2.8$ compared to that of the blue dashed line $q_{\rm out}=1/2$, the boundary of the semi-secular regime of Zeta Aquarii is below the blue dashed line, bringing Zeta Aquarii further away from the semi-secular regime. 
 Our adopted stability limit is not very sensitive to the outer mass ratio, but it is affected by the inclination of the system. As Zeta Aquarii has a retrograde orbit with an inclination of $140^{\circ}\pm 10^{\circ}$ \citep{Tok16b} it is relatively more stable than a prograde equivalent system. The stability limit of Zeta Aquarii  lies below the solid red line in Fig.\,\ref{fig:theory_p}. Ultimately, the proximity of Zeta Aquarii makes it an interesting system in the analysis of stability limits and strong three-body effects. With an age of about 3Gyr and LK cycles with timescales of about 48000 yr, the system has likely gone through many LK oscillations as corroborated  by the inclination and the high inner eccentricity of 0.87 \citep{Tok16b}. 

In the future when new triple systems are discovered, Figs.\,\ref{fig:theory}~and~\ref{fig:theory_p} can be used to assess which processes are relevant for that given triple. An interactive version of Fig.\,\ref{fig:theory}a is available at \href{https://bndr.it/wr64f}{\color{blue}https://bndr.it/wr64f}.
The graphical interface allows the user to adjust the figure to any given triple.

  \begin{figure*}
    \centering
    \begin{tabular}{cccc}
 \multicolumn{4}{c}{ \includegraphics[clip=true, trim =0mm 60mm 0mm 60mm, width=0.75\textwidth]{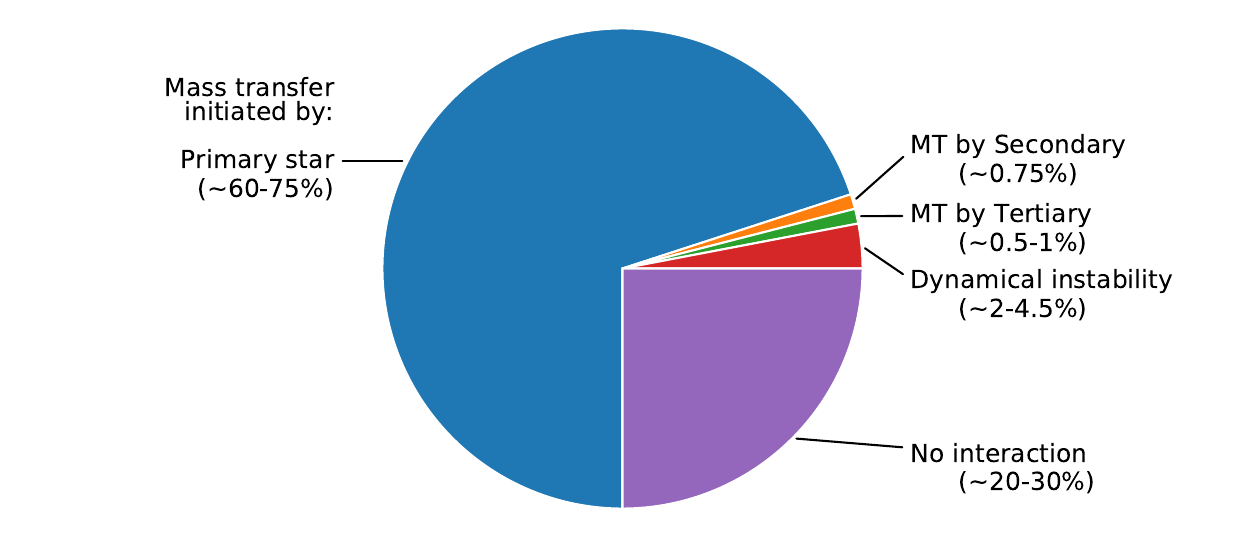} } \\
Model OBin & Model T14 & Model E09 & \\
\includegraphics[clip=true, trim =60mm 60mm 60mm 60mm, width=0.2\textwidth]{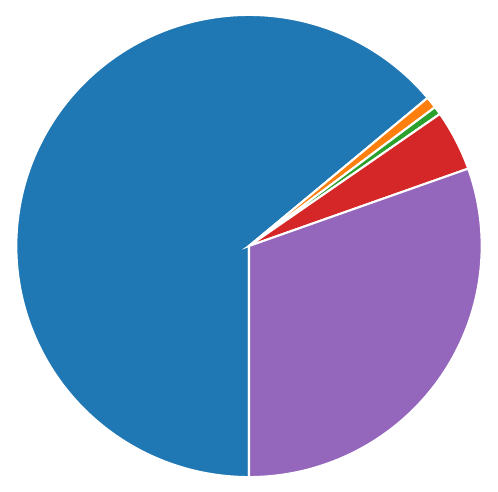}  &
 \includegraphics[clip=true, trim =60mm 60mm 60mm 60mm, width=0.2\textwidth]{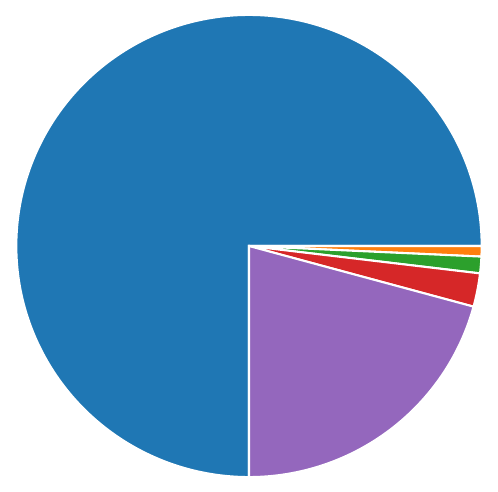} & 
 \includegraphics[clip=true, trim =60mm 60mm 60mm 60mm, width=0.2\textwidth]{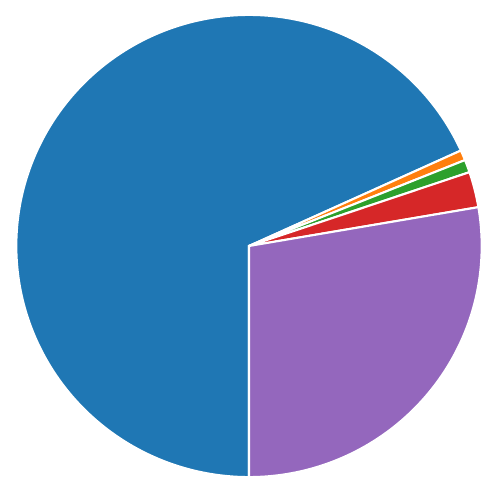} &
\includegraphics[clip=true, trim =0mm 155mm 149mm 0mm, width=0.2\textwidth]{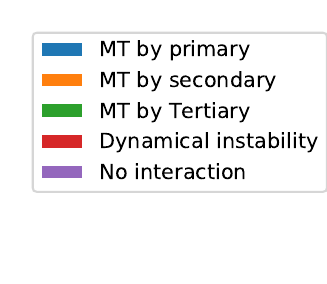}  \\
	\end{tabular}
    \caption{Frequencies of the most abundant evolutionary channels for triples with intermediate-mass primaries based on a population synthesis simulation of triple systems using \code. Mass transfer is common to occur in the evolution of a triple, and primarily with the primary star of the inner binary donating mass. On the top a summary of all models are shown indicating the range in expected frequencies.  On the bottom row pie charts are shown for the different models individually. }
    \label{fig:pie}
    \end{figure*}

\section{Triple evolutionary pathways} 
\label{sec:res}
Based on the models described in Sect.\,\ref{sec:init_distr_tr}, the expected birthrate of all triples is about 150-175 systems per 1000\Msolar\, of created stars (Tbl.\,\ref{tbl:birthrates}). In the Milky Way, this translates to a birthrate of about one system every two years. As expected from the IMF, the majority of triples have low-mass stellar companions that do not leave the MS in a Hubble time. About 10\% of triples have an intermediate mass primary (i.e. the initially most massive star in the inner binary), and $\sim$1-2\% has a high-mass primary which are the typical neutron star and black hole progenitors. In this study we focus on triple systems with intermediate mass primaries. The evolution of these triple star systems  can be split up into five distinct evolutionary pathways:
\begin{itemize}
\item The primary star initiates mass transfer;
\item The secondary star initiates mass transfer ;
\item The tertiary star initiates mass transfer;
\item The triple becomes dynamically unstable;
\item The stars do not interact and the system remains intact for a Hubble time.
\end{itemize}

 Fig.\,\ref{fig:pie} shows the average frequency of each channel based on our triple population synthesis simulations. 
For 70-80\% of triples, the stars interact with each other at some point during their evolution. 
The most common outcome of triple evolution is mass transfer in the inner binary initiated by the primary. A few percent of triples destabilise or experience a mass transfer phase initiated by the secondary or tertiary, which all three is not expected to occur in isolated binaries without tertiary companions.

The birthrates of the five channels do not depend strongly 
on our assumptions for the primordial populations of triples (Tbl.\,\ref{tbl:birthrates}). The rates vary within a factor of about two between the different models. 
In the subsequent sections, we discuss in more detail the characteristics of each channel.

\subsection{Mass transfer initiated by the primary star}
 \label{sec:mt1}
 
\subsubsection{In isolated triples } 
 
The most likely outcome for a triple with a $1-7.5\Mo$ mass primary is for the primary to fill its  Roche lobe equivalent. This happens in  64-73\% of all triples depending on the modelling of the primordial population.

\begin{figure}
    \centering
	\includegraphics[width=\columnwidth]{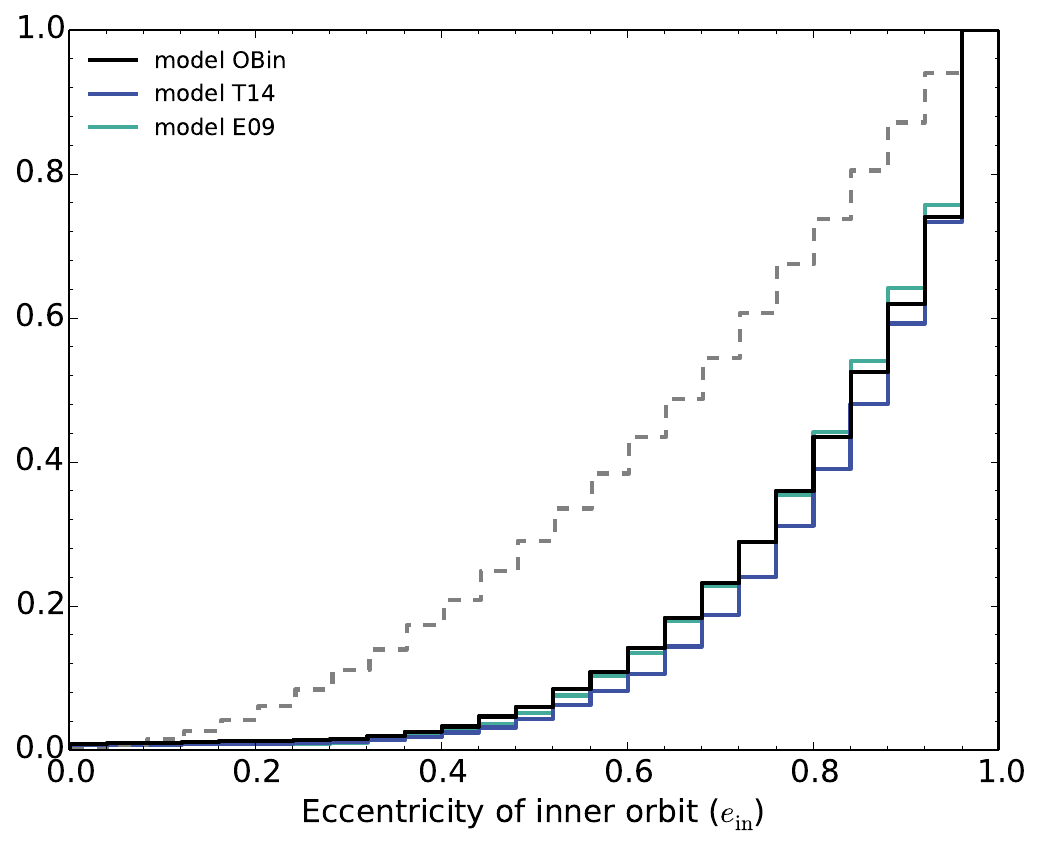} \\
    \caption{Cumulative eccentricity distribution at the onset of mass transfer for the three different models normalised to unity (solid lines). The dashed lines represent the initial eccentricity distribution at ZAMS. The solid lines reflect those systems that are eccentric upon the first RLOF, which is about 40\% of systems.} 
    \label{fig:mt1_ecc_hist}
    \end{figure}

\begin{figure}
    \centering
	\includegraphics[width=\columnwidth]{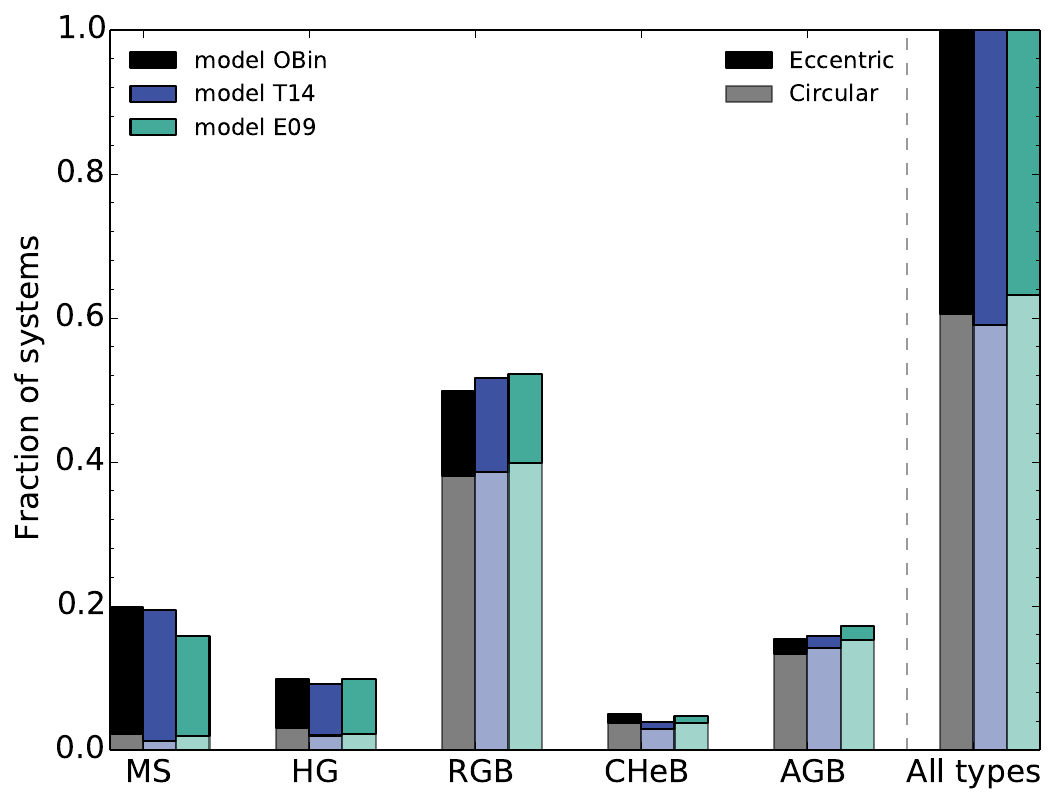} 
    \caption{Evolutionary states of the donor stars at the onset of mass transfer in circularised as well as eccentric orbits. The fractions are given with respect to the total number of systems where the primary fills its Roche lobe. Non-evolved donors predominantly initiate mass transfer in eccentric orbits and vice versa. } 
    \label{fig:mt1_donor_hist}
    \end{figure}

Whereas isolated binaries are typically expected to circularise before (and remain circular during) the mass transfer phase
\footnote{An exception occurs in massive binaries. \cite{Eld09} show that not all binaries with massive donor stars (and radiative envelopes) circularise during Case-B mass transfer events if the mass transfer occurs on a timescale that is shorter than what is required to circularise the orbit by tides. },
this is not the case for triple stars. An appreciable fraction of the inner orbits of triples is still eccentric at the onset of mass transfer (Fig.\,\ref{fig:mt1_ecc_hist}). The distribution of eccentricities of the inner orbits is bimodal; orbits are either circularised or highly eccentric. In the latter case, the median eccentricity is 0.86-0.89. Of the inner orbits 59-63\%  have circularised and 37-41\% is eccentric (Fig.\,\ref{fig:mt1_donor_hist}).

Whether or not an orbit circularises before the onset of mass transfer is a trade-off between tides and the LK cycles. 
The LK cycles can increase and decrease the eccentricity on a timescale that is strongly dependent on the periods of the inner and outer orbit (Eq.\,\ref{eq:t_kozai}). In the absence of LK cycles, such as for isolated binaries, tides act to circularise the orbit. 
The characteristic timescale for orbital change due to tides is strongly dependent on the type of star and the ratio of the stellar radius to its orbital separation \citep{Hut80}. The interaction with tides and LK cycles, is one example of the interplay  between stellar evolution and three-body dynamics in the evolution of stellar triples. The eccentricity distribution upon mass transfer demonstrates that three-body dynamics affects the evolution of a significant fraction of all triples.

\begin{figure}
    \centering
    \begin{tabular}{c}
	\includegraphics[clip=true, trim =0mm 15.5mm 0mm 0mm,width=\columnwidth]{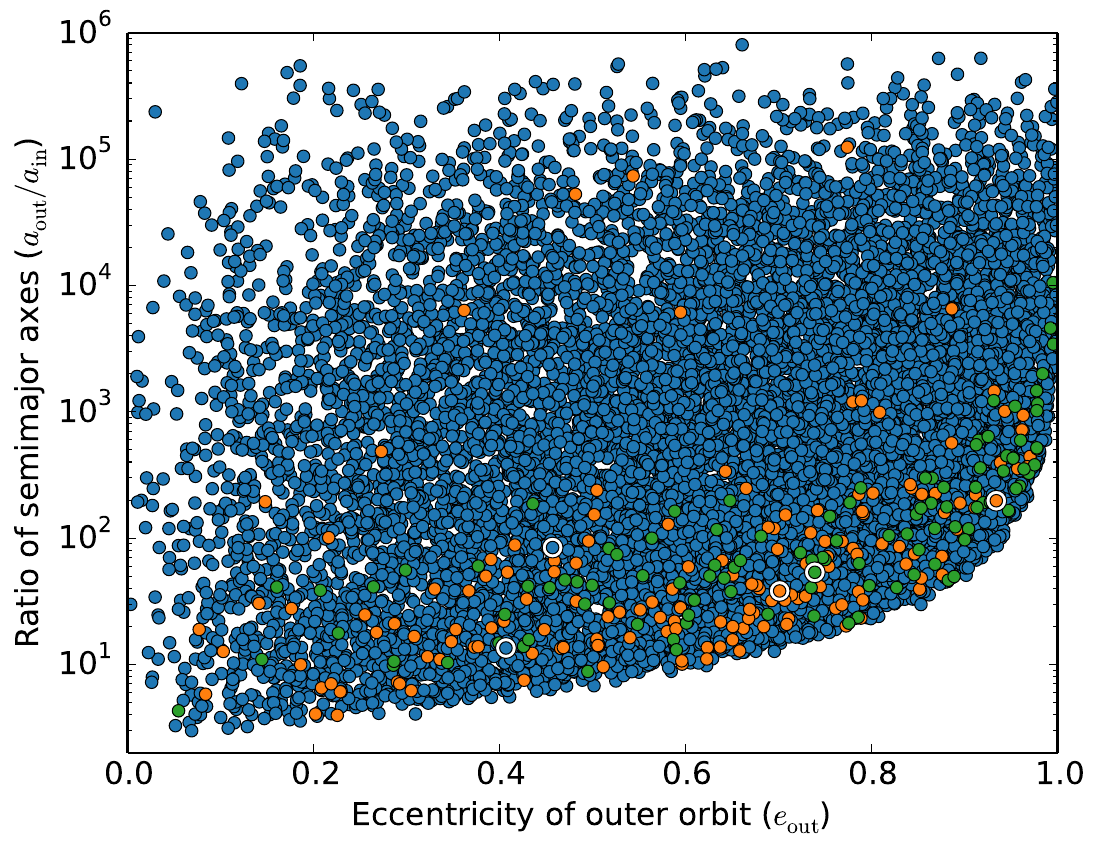} \\[-1mm]
	\includegraphics[clip=true, trim =0mm 0mm 0mm 4mm,width=\columnwidth]{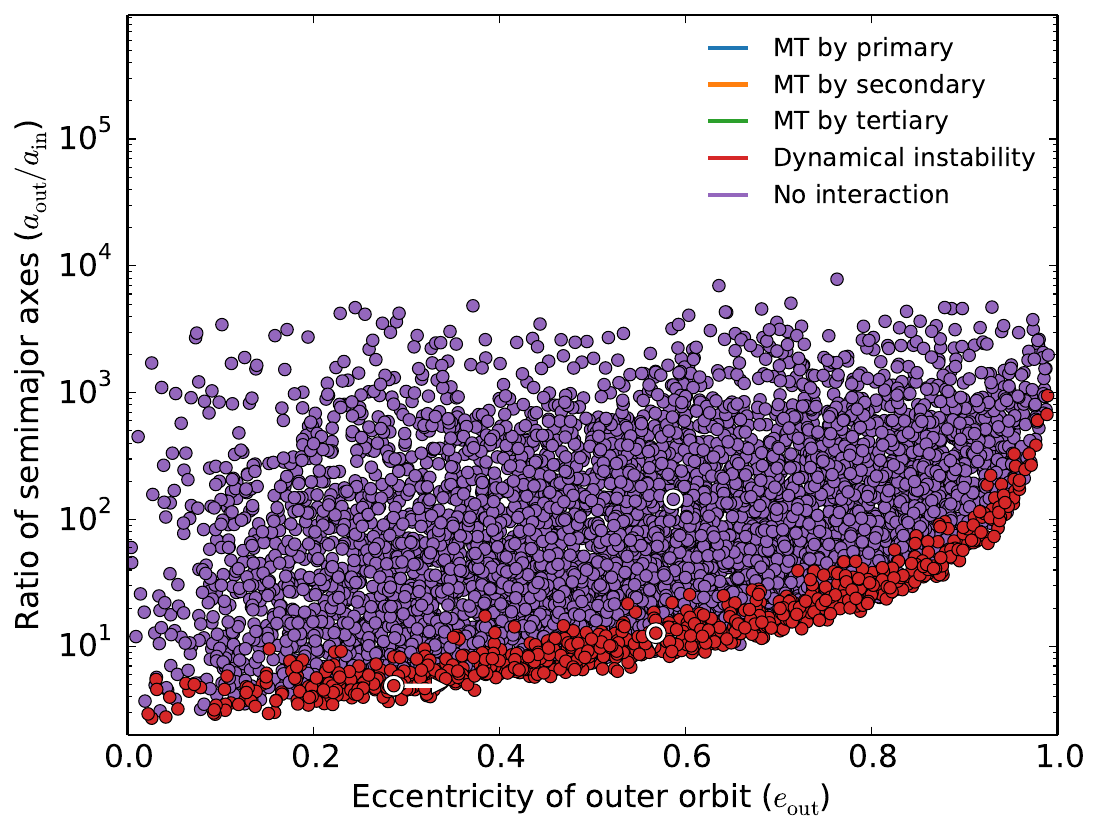} \\
	\end{tabular}
    \caption{ Visualisation of the initial population of triples of the default model OBin, which highlights the different dynamical regimes as shown in Fig.\,\ref{fig:theory}. The examplary systems mentioned in the text are marked with white halos. } 
    \label{fig:theory_data}
    \end{figure}

An example of such an evolution is given by the following system: 
$m_1=1.63\Mo$, $m_2=0.82\Mo$, $m_3=2.06\Mo$,   $i=71.5^{\circ}$,  $a_{\rm in}=8.91\Ro$,  $e_{\rm in}=0.57$, $g_{\rm in} = -2.53 $rad, $a_{\rm out} =750\Ro$, $e_{\rm out}= 0.46$, $g_{\rm out}= 1.55$rad on the ZAMS. 
Initially the system is detached, dynamically stable, and in the octupole regime with $\epsilon_{\rm oct} = 0.0023$, see also Figs.\,\ref{fig:theory_data}. 
Correspondingly the timescale of the Lidov-Kozai cycles is short, that is about 2680 years. 
After about 50 Myrs, when the inner eccentricity has increased to 0.61 and the Roche lobe of the primary has shrunken accordingly, the primary star fills its Roche lobe and initiates the mass transfer phase in a significantly eccentric orbit. 
Similar to this example, it is common for all systems of this evolutionary channel to be in the octupole regime, as can be seen by comparing Fig.\,\ref{fig:theory_data} and Fig.\,\ref{fig:theory}. 
In model OBin and model E09 it entails about 45$\%$ (14$\%$) of systems that have $\epsilon_{\rm oct}>0.001\, (0.01)$. 
The fraction increase to 61$\%$ (21$\%$) respectively for model T14, as the inner and outer orbits tend to be closer together in this model. 
Fig.\,\ref{fig:theory_data} also shows that the systems of this evolutionary channel are not only in the octupole regime at birth, but cover a range of dynamical regimes, which indicates a variable importance of the tertiary star and three-body dynamics.

The eccentricity is relevant for the subsequent evolution of the system; the mass transfer phase progresses differently when the orbit is eccentric instead of circularised due to the phase-dependency of the mass transfer when the orbit is eccentric  \citep{Lay98, Reg05, Laj11, Hel16, Bob17, Ham19}.
Whereas tides act to circularise the orbit, enhanced mass loss at (or close to) periastron (either from stellar winds or RLOF) can enhance the eccentricity, such that some binaries can remain eccentric for long periods of time \citep{Win95, Sok00, Sep07b, Bon08, Sep09, Vos15, Dou16b, Kas18}. Mass transfer in eccentric orbits is observationally supported by a number of eccentric semidetached and contact systems \citep{Pet99, Bof14}, systems close to RLOF \citep{Gua04, Rag05, Bon08, Nic12,Wal15}.

 Fig.\,\ref{fig:mt1_donor_hist} shows the evolutionary phase of the donor star at the onset of mass transfer. 
 The most common donor stars are on the red giant branch (RGB, 50-52\%), followed by the main-sequence (MS, 16-20\%) and the asymptotic giant branch (AGB, 15-17\%). 
Moreover, Fig.\,\ref{fig:mt1_donor_hist} shows that mass transfer in an eccentric orbit is most likely to occur with a fairly unevolved donor star (MS-RGB) compared to stars in the core helium burning phase (CHeB) or on the AGB. In particular for donors on the MS and Hertzsprung-gap (HG) that have weak tides, the inner orbits are typically still eccentric upon mass transfer; in roughly 9 out of 10 cases.   
   
These results also depend on the mass range of the donor star. 
For systems with relatively high primary masses, it is relatively more common that the donor star is on the MS or HG gap upon Roche lobe overflow, and less common to be on the RGB. For example, while over $40\%$ of MS donors have a high-mass origin (here $m_1>2\Mo$), less than 15\% of RGB donors are massive. 
Additionally, the eccentricity distribution is affected as well, because the tidal timescale of the star is linked to its mass and evolutionary state. 
For example, more massive MS stars have shorter tidal timescales, and therefore a larger fraction of systems are circularised upon mass transfer, that is 12\% for $m_1>2\Mo$ versus 8.5\% for $1.25\Mo<m_1<2\Mo$ for MS donors in model OBin. Even lower mass stars have convective envelopes and corresponding even shorter tidal timescales, such that 16\% of systems are circularised.  We note however, that the overall trend that mass transfer in an eccentric orbit is most likely to occur with a fairly unevolved donor star is robust against variations in the mass range in our simulations.

\begin{figure}
    \centering
	\includegraphics[width=\columnwidth]{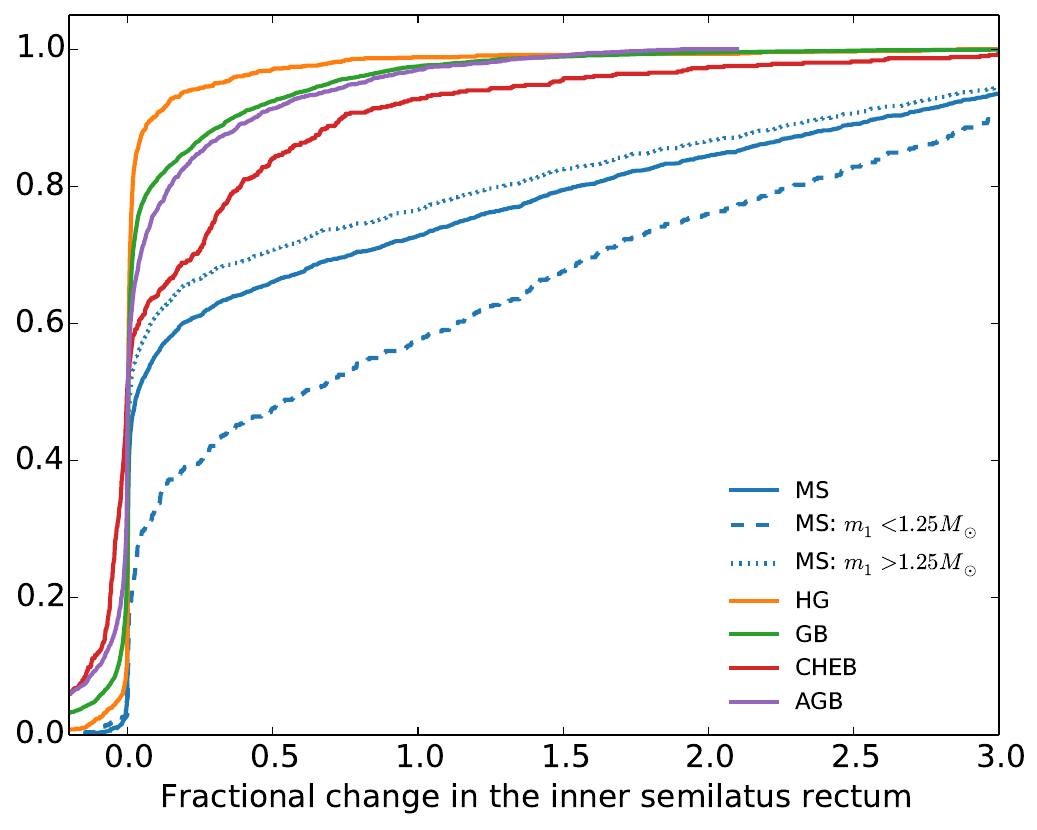} 
   \caption{Change in the semilatus rectum of the inner orbit ($a_{\rm p}\equiv a_{\rm in}(1-e^2_{\rm in})$) between the zero-age MS and the onset of mass transfer. The figure shows the cumulative distribution of the logarithm of $\mathcal{A} = a_{\rm p, init}/a_{\rm p, final}$ for model OBin, while other models show qualitatively similar behaviour.  LK cycles with tidal friction drive the inner stars closer together ($\mathcal{A}>1$), whereas stellar winds act to widen the orbit ($\mathcal{A}<1$). The figure shows that LKCTF can shrink the inner orbit most efficiently for low-mass MS stars, followed by higher-mass MS stars, and evolved stars.  }
    \label{fig:LKCTF}
    \end{figure}

The interplay of LK cycles with tidal friction (LKCTF) can give rise to stellar migration \citep{Maz79,Kis98,Fab07, Nao14, Liu15, Bat18, Ham19d}.
During the high-eccentricity parts of the LK cycles, the inner stars pass each other at short periastron distances when tidal dissipation is very efficient. 
At this point the tides act to reduce the eccentricity as well as the semi-major axis of the inner orbit. LKCTF has been studied extensively for low-mass MS stars  (for stellar triples as well as planetary systems), but a test of the efficiency for evolved stars is missing so far. 
We aim to quantify the importance of stellar migration for different stellar donors by studying the change in the semilatus rectum of the inner orbit ($a_{\rm p}\equiv a_{\rm in}(1-e^2_{\rm in})$) from the initial to the final orbit, that is  $\mathcal{A} \equiv a_{\rm p, init}/a_{\rm p, final}$  in Fig.\,\ref{fig:LKCTF}. 
If a system would suffer from strong tidal circularisation (without stellar winds and LK oscillations), $\mathcal{A} = 1$. If instead the stellar wind mass losses are substantial, $\mathcal{A} < 1$, such as during the giant stages. If instead stellar migration occurs due to LKCTF, $\mathcal{A} > 1$.  Fig.\,\ref{fig:LKCTF} shows a clear difference between different evolutionary states of the donor stars. Over 75\% of systems with low-mass MS donors experience stellar migration (here $\mathcal{A}>1.05$), whereas this is over 40\% for their higher-mass counterparts, and over 30\% for evolved donors. The amount of orbital shrinkage ranges from a factor of a few for systems with evolved donors to several orders of magnitude for donors on the MS.

The delay times between the onset of the mass transfer and the formation of the triple range from several hundreds of Myr to many Gyr. 
We find a few percent of systems ($\lesssim 5\%$) 
that experience RLOF within 1Myr in this channel. For  less than $\sim$1\% the mass transfer starts within 0.01Myr. These systems have strong LK cycles with comparable timescales and an inclination distribution that peaks around 90$^{\circ}$. Although technically the systems are detached and dynamically stable at initiation according to Eq.\,\ref{eq:stab_crit}, it is questionable whether these systems could have formed in the first place; very likely the stars would have had contact or strong tidal evolution during the pre-MS phase when the radii of the stars were larger.

\subsubsection{Comparison to isolated binaries}
In this section we study the role of the tertiary star for the systems where the primary initiates mass transfer. We study how these systems would evolve without the presence of the tertiary star. We take the inner binaries from model OBin, T14 and E09, remove the tertiary, and assume their orbits completely circularise before RLOF due to tides to the semilatus rectum. The stellar evolution, including winds and its effect on the orbits, is taken into account in the same way as in \code.

      \begin{figure}
	\includegraphics[width=\columnwidth]{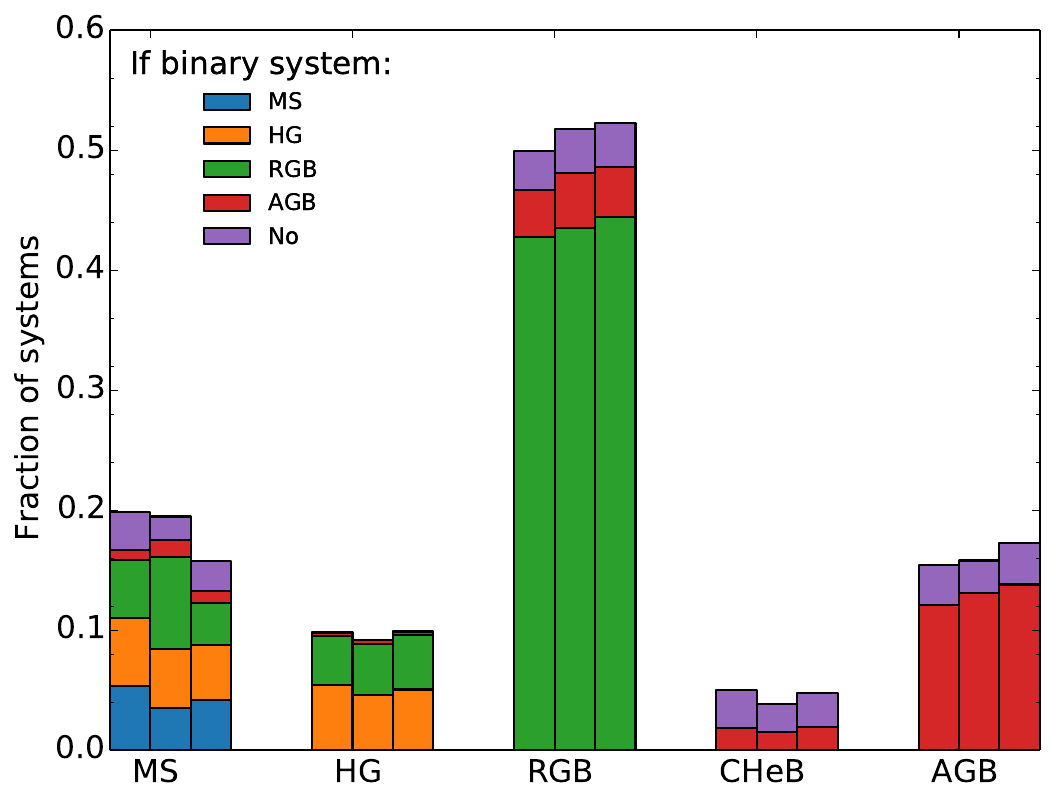} 
    \caption{Evolutionary states of the donor stars at the onset of mass transfer. For each type of donor star, three bars are shown, with the left representing model OBin, middle model T14, and on the right model E09, similar to Fig.\,\ref{fig:mt1_donor_hist}.  With colour we indicate when the donor star would initiate mass transfer if the inner binary would be isolated: from the bottom to the top, on the MS (blue), HG (yellow), RGB (green) or AGB (red), or if mass transfer happens at all (if not: purple). Overall, the onset of RLOF is expedited due to the presence of the tertiary star and the frequency of case A mass transfer increases by a factor 3.7-5.5. } 
    \label{fig:mt1_donor_hist_bin}
    \end{figure}

Overall, we find that for the same binary, mass transfer occurs earlier in its evolution if the binary is orbited by a third body.
As an example we mention the following system: $m_1=1.33\Mo$, $m_2=1.07\Mo$, $m_3=0.26\Mo$,   $i=71.5^{\circ}$,  $a_{\rm in}=70.1\Ro$,  $e_{\rm in}=0.65$, $g_{\rm in} = -0.38$rad, $a_{\rm out} =945\Ro$, $e_{\rm out}= 0.41$, $g_{\rm out}= -0.88$rad on the ZAMS.  The system undergoes strong LK cycles with timescales of the order of a thousand years.  The system is even in the eccentric LK regime as indicated by an octupole parameter of $\epsilon_{\rm oct}\approx 0.004$, and its location in Fig.\,\ref{fig:theory_data}. 
As the inner eccentricity increases from 0.65 to 0.94, the primary star fills its Roche lobe. This happens after 2.8Gyr when the primary star is still on the MS. Without the presence of the tertiary star, the eccentricity of the (previously inner) orbit would not increase, but rather decrease until circularisation is reached. At that point the orbital separation is about 40\Rsolar. This is too large for the system to experience mass transfer when the primary star is still on the MS. However afterwards, when the primary star has crossed the Hertzsprung gap and is ascending the giant branch, the primary star finally fills its Roche lobe.

Fig.\,\ref{fig:mt1_donor_hist_bin} shows how often mass transfer is brought forwards due to the presence of the tertiary star. For example,  in 9.5-11\% of isolated inner binaries the primary fills its Roche lobe on the Hertzsprung gap. If the tertiary is taken into account, 47-52\% of these systems already interacts when the donor star is still on the MS.  
However, the total fraction of HG donors in triples (9-10\%) is roughly similar to that in binary systems, as binary systems with more evolved-donors (mostly on the RGB) experience mass transfer earlier when in a triple. Secondly, we note that in 3.9-5.0\% of triples the primary star is a core-helium burning star at the onset of the mass transfer, whereas in intermediate-mass binaries we do not anticipate this to occur. The difference between binaries and triples is largest for MS donors. Whereas in intermediate-mass binaries, merely 3.5-5.3\% experience case A mass transfer \citep{Kip90}, for triples this increases to 15.8-19.9\%. The majority of case A mass transfer in triples is brought about by three-body dynamics, which is also supported by their eccentricity distribution (Fig.\,\ref{fig:mt1_donor_hist}).

In the experiment of removing the tertiary star from the inner binary, we also find that some inner binaries do not experience mass transfer without the tertiary. This is the case for 11-13\% of the triples, with no preference for the evolutionary state of the donor star (Fig.\,\ref{fig:mt1_donor_hist_bin}). For the full populations of triples (not only the ones where the primary initiates a mass transfer phase), 65-77\% of triples experience mass 
 transfer, while this decreases to 58-70\% when removing the tertiary. In conclusion, mass transfer can occur in 10\% more systems in a triple population than if the inner binaries were isolated. 
 
 However, realistic isolated binaries have different orbits than the inner binaries of triples (see bottom panel of Fig.\,\ref{fig:init_a}). To assess the frequency of mass transfer in binaries and the types of their donor stars, we construct two sets of initial binary populations. The models differ with respect to the distribution of orbital separations; an uniform distribution in the logarithm of the orbital separation between 5\Rsolar\,to $5\cdot 10^6$\Rsolar\, (green dash-dotted curve in Fig.\,\ref{fig:init_a}), or the log-normal distribution in periods described in Sect.\,\ref{sec:init_distr_tr} (black dotted curve in Fig.\,\ref{fig:init_a}). The former and latter model are related to model OBin and T14 respectively, in which both $a_{\rm in}$ and $a_{\rm out}$ are drawn from the same distribution, but the requirement of dynamical stability skews the orbital distributions (Sect.\,\ref{sec:init_distr_tr}). In our binary models the primary mass, secondary mass and eccentricity are distributed in a similar fashion as $m_1$, $m_2$, and $e_{\rm in}$. 
  In the population with the log-uniform distribution we find RLOF can take place in only 39\% of systems, compared to 65\% of triples in model OBin. Adopting the log-normal distribution, the fraction drops even further to 28\%, whereas mass transfer occurs in 77\% of triples in model T14. In conclusion, the tertiary star can drive inner binaries to RLOF, that would not have interacted without the tertiary present. This is partly because of the LK-cycles the tertiary can raise on the inner binary, but predominantly because the inner binaries are typically more compact than isolated binaries in order for the triple to be dynamically stable.

 \begin{figure*}
    \centering
    \begin{tabular}{ccc}
	\includegraphics[width=0.32\textwidth]{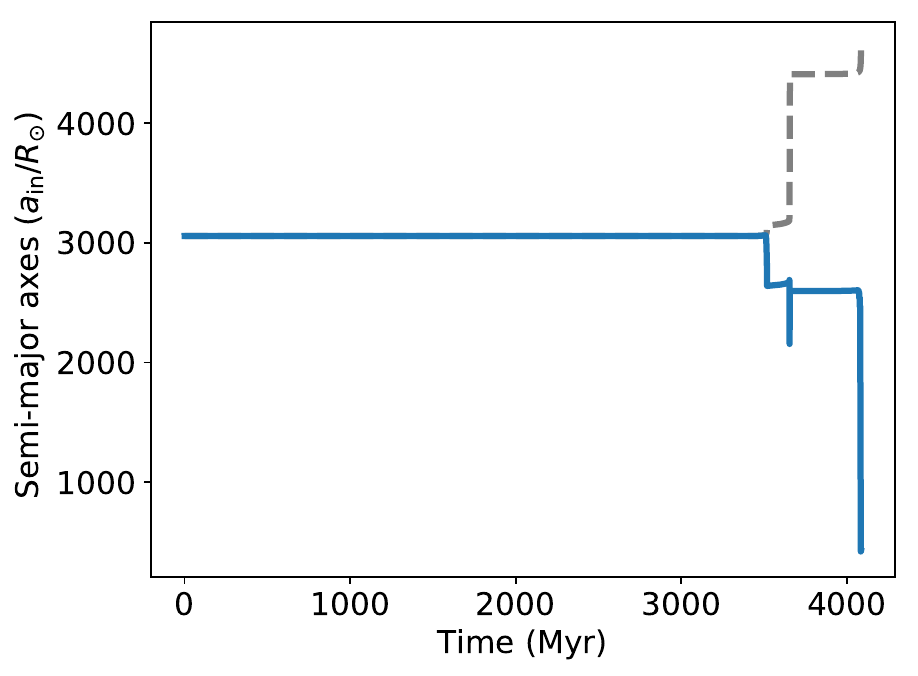} 
	\includegraphics[width=0.32\textwidth]{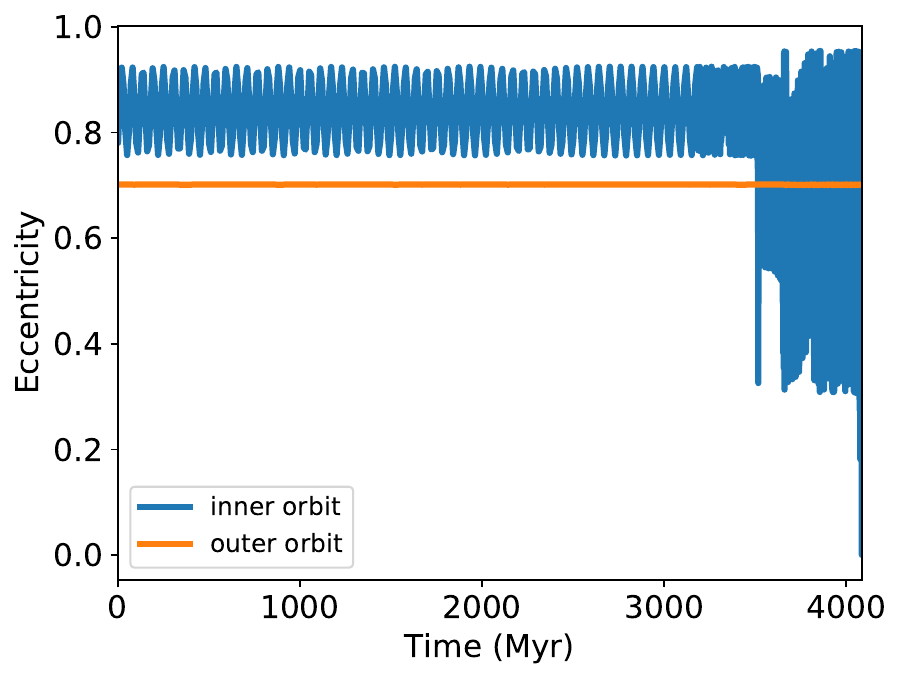} 
	\includegraphics[width=0.32\textwidth]{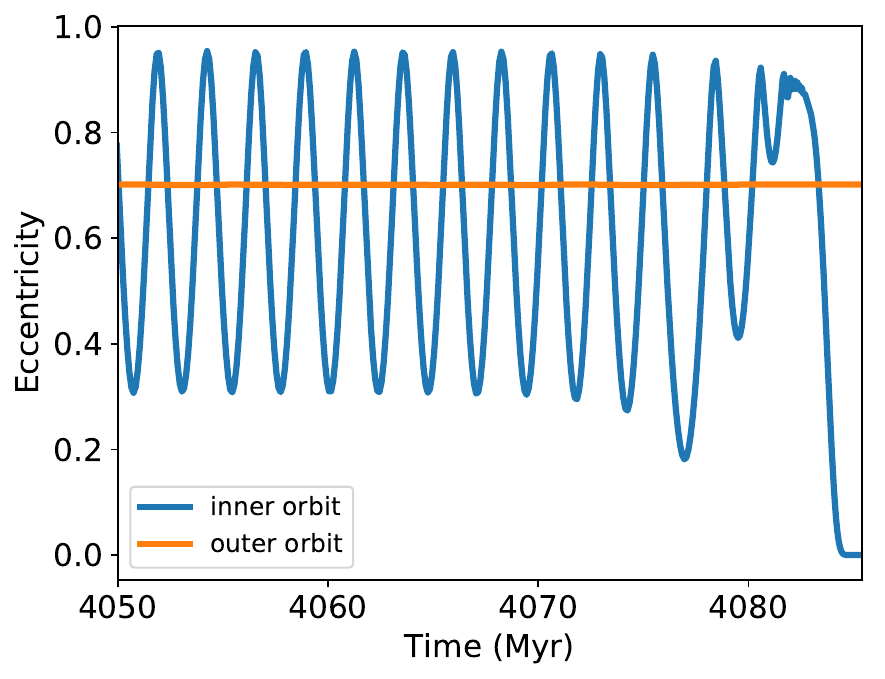} 
	\end{tabular}
    \caption{First example of a system in which the secondary initiates the first mass transfer phase while the primary has already evolved to the WD stage. In this case the interplay between LK cycles, stellar evolution and tides shrinks the inner orbit in consecutive phases; when the primary evolves along the RGB at $\sim$ 3500 Myr, on the AGB at $\sim$3650 Myr, and when the secondary is on the RGB at $\sim$4100 Myr. The top and middle panel show the inner semimajor axis and eccentricity as a function of time. The bottom panel shows the circularisation of the inner orbit in the final moments before the mass transfer. In the top panel we have overplotted in dashed grey the widening of the inner orbit if only stellar winds would act upon it. The initial conditions of this system are: $m_1=1.42\Mo$, $m_2=1.36\Mo$, $m_3=0.70\Mo$,   $i=70.9^{\circ}$,  $a_{\rm in}=3057\Ro$,  $e_{\rm in}=0.86$, $g_{\rm in} = 1.30 $rad, $a_{\rm out} =116551\Ro$, $e_{\rm out}= 0.70$, $g_{\rm out}= -1.99$rad.  
     } 
    \label{fig:mt2_ex_wss_ll_el}
    \end{figure*}

\subsection{Mass transfer initiated by the secondary star}
  \label{sec:mt2}
The second evolutionary channel involves mass transfer in the inner binary where the secondary star is the donor star. In contrast, in the previous section it was the primary star (i.e. the initially most massive star) that initiated the mass transfer event. In our simulations, 0.7-0.8\% of triples experience mass transfer initiated by the secondary star. This gives a Galactic event rate of about 3-4 times every $10^4$ years. We differentiate between two scenarios; those where the primary star is still experiencing nuclear burning and those where the primary has become a WD at the onset of the mass transfer. The latter is slightly more common, comprising $53-61$\% of the systems.

In the first scenario typically both the primary and the secondary are on the MS. Their masses are similar with a median mass ratio of more than 0.93. Though the secondary fills its Roche lobe, the primary is very close to Roche lobe overflow as well. Typically the radius of the primary is within a few percent of the associated Roche lobe. The system likely becomes a contact binary such that the difference between the primary and secondary star is an academic one. Additionally, there are systems for which the mass transfer starts within $\sim$1Myr since formation. As discussed in  Sect.\,\ref{sec:mt1}, it is likely these systems would have interacted previously on the pre-MS. It comprises 42\% of the systems in the first scenario. The fraction goes up to 61\% for model T14.

     \begin{figure*}
    \centering
    \begin{tabular}{cc}
	\includegraphics[width=0.8\columnwidth]{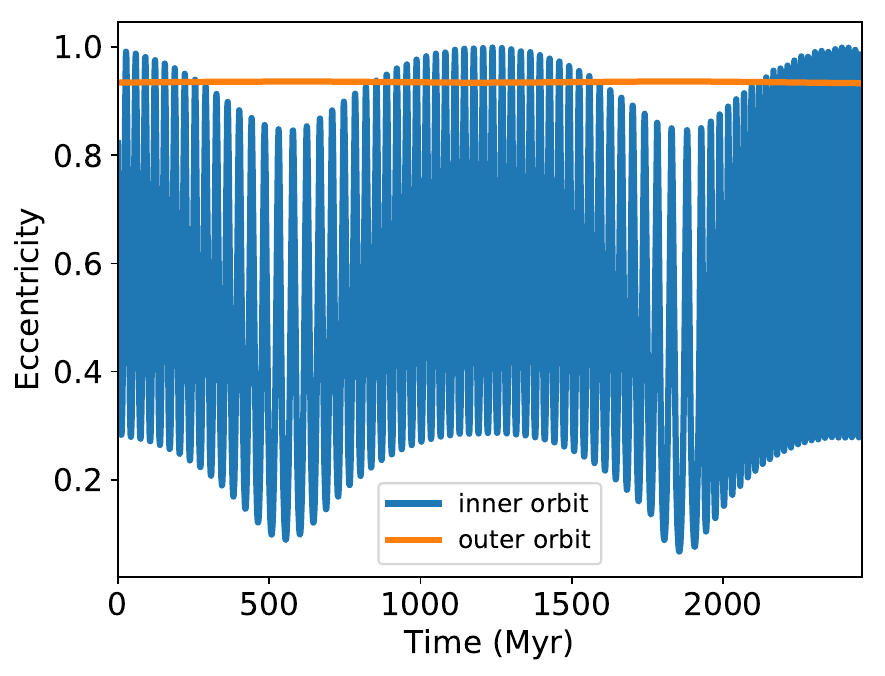} &
	\includegraphics[width=0.8\columnwidth]{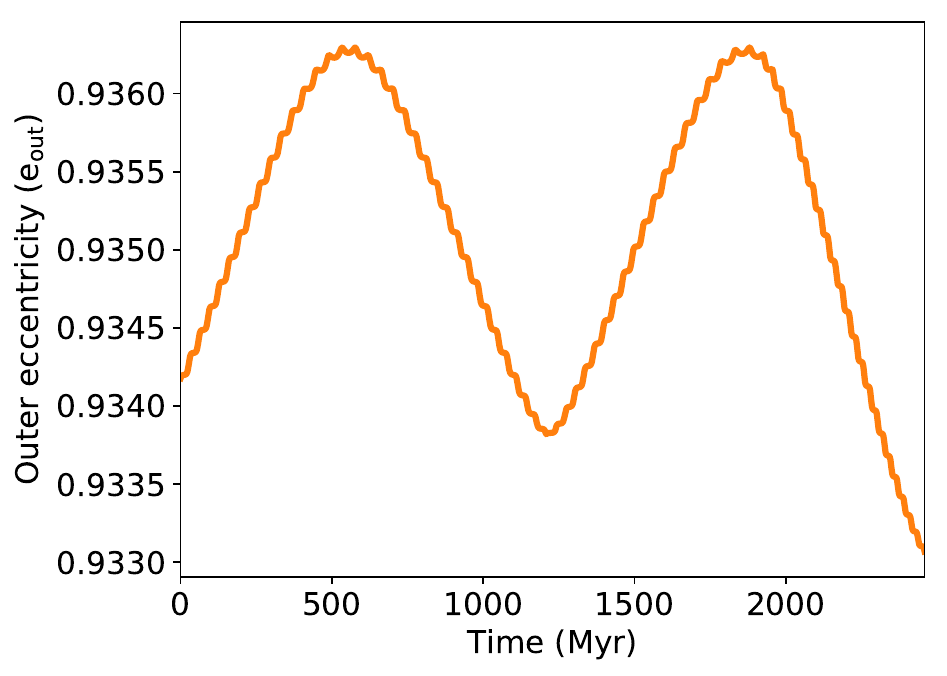} \\
	\includegraphics[width=0.8\columnwidth]{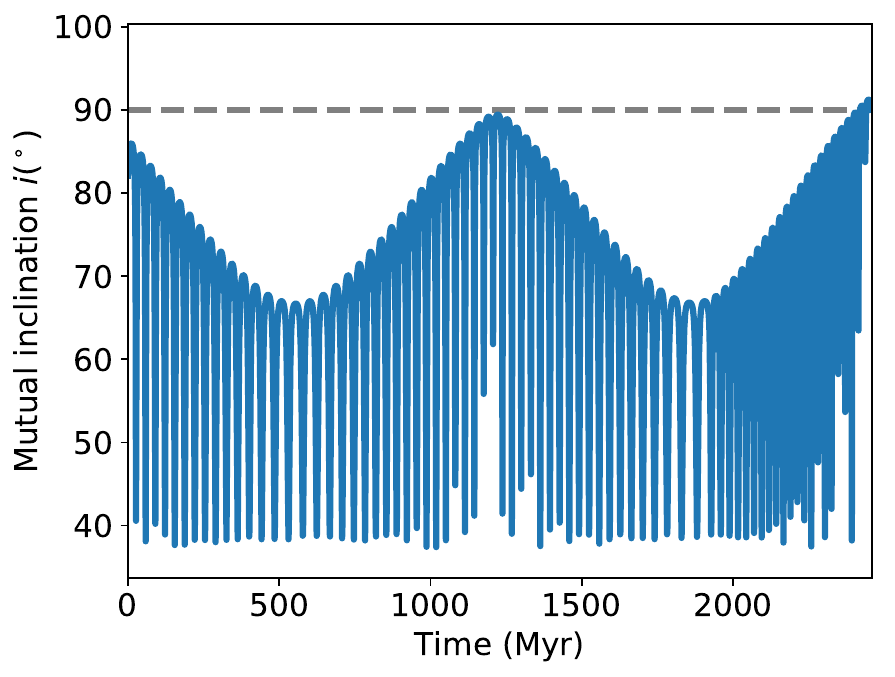} &
	\includegraphics[width=0.8\columnwidth]{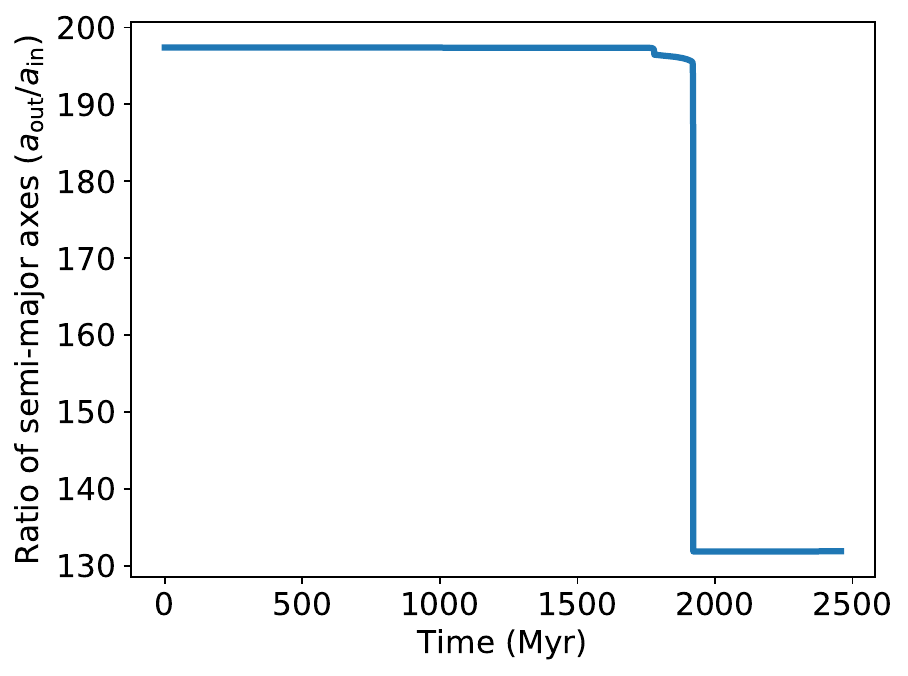} \\
	\end{tabular}
    \caption{Second example of a system in which the secondary initiates the first mass transfer phase while the primary has already evolved to the WD stages. Here the interplay between the different terms of the three-body dynamics effects and stellar evolution makes the inner orbit extremely eccentric, leading to the mass transfer phase. The top two panels show the eccentricities as a function of time, followed by the inclination and the ratio of the outer to inner semimajor axis as a function of time. 
    The short and long term modulations seen in these figures represent the timescales of the quadrupole and octupole terms, respectively. After the primary has become a WD at $\sim$1900Myr, and the inner and outer orbit have approached each other, the system experiences a flip in the orbital inclination (when crossing the black dashed line of $i=90^{\circ}$), accompanied by an extreme eccentricity in the inner orbit and a subsequent mass transfer event.  The initial conditions of this system are: $m_1 = 1.76\Mo$, $m_2 = 0.26\Mo$, $m_3=1.22\Mo$, $i=82.1^{\circ}$, $a_{\rm in} =7123\Ro$, $e_{\rm in}=0.82$,  $g_{\rm in}= -0.63$rad, $a_{\rm out} = 1406013\Ro$, $e_{\rm out} = 0.93$, $g_{\rm out}= -1.12$rad. 
} 
    \label{fig:mt2_ex_wss_hh}
    \end{figure*}

In the second scenario  the primary star has already evolved to the remnant stages without initiating mass transfer. 
When the secondary star fills its Roche lobe, mass is transferred to a compact object, such that cataclysmic variables and low-mass X-ray binaries could be formed in this way. Typically the mass transfer phase starts
 at late times that is several Gyrs after star formation. 
 If the inner binary were to be isolated, the evolution of this second scenario would not be possible: if the orbit is wide enough such that the primary does not fill its Roche lobe during its evolution, the secondary star does not either\footnote{This is strictly only true for low and intermediate mass stars, as for these stars the maximum  radius attainable during the stellar life increases with the initial mass}.
The second scenario is possible for triples due to the interplay of three-body dynamics with other physical processes. Overall, this occurs in two different ways which are equally likely to occur:\\
- In one case, the inner orbit shrinks due to the interplay of LK cycles with tidal friction (i.e. LKCTF) as shown in Fig.\,\ref{fig:mt2_ex_wss_ll_el}. When the primary star is in the evolved stages of evolution, and close to RLOF, tides are efficient to circularise the inner orbit as well as reduce its semimajor axis,  for example  at $\sim$ 3500 Myr and $\sim$3650 Myr in Fig.\,\ref{fig:mt2_ex_wss_ll_el}. The contraction is counteracted by the mass loss in the stellar winds of the primary star (as seen at $\sim$3650Myr). After the primary has become a WD, and the secondary star evolves, tides make the inner orbit contracts for the third time, and circularise. Ultimately this leads to mass transfer from the secondary star to the (primary) WD component  at $\sim$4100Myr.
In this channel the secondary stars are typically evolved stars from the RGB to the AGB. For roughly half of the systems, the inner orbits reach complete circularisation. \\
- A different pathway of evolution is shown in Fig.\,\ref{fig:mt2_ex_wss_hh}. For this type of systems, the inner orbits have widened in response to the wind mass losses of the primary stars. The orbits are extremely eccentric at the moment of RLOF, on average $e_{\rm in} \gtrsim 0.99$. At periastron, the stars pass at each other closely, but do not collide directly.
The secondary stars are typically low-mass ($m_2\lesssim 1\Mo$) and  on the MS. 
This type of system shows a rich variety of three-body dynamics. The octupole term of the three-body dynamics is relevant (i.e. $|\epsilon_{\rm oct}| \gtrsim 0.001$) from the birth of the systems until the onset of the mass transfer\footnote{This is in contrast with the previous pathway in the second scenario. For those systems the octupole parameter has a median value of $(1-3)\cdot 10^{-4}$  initially and $(3-5)\cdot 10^{-4}$ at the onset of mass transfer, which is significantly below where the octupole terms are expected to be important.}.  
An example evolution is shown in Fig.\,\ref{fig:mt2_ex_wss_hh}. It shows distinct modulations in the inner eccentricity and inclination both on short and long timescales that are linked to the quadrupole and octupole term, respectively.  As the primary star evolves off of the MS, it loses mass in its stellar wind, most strongly at an age of 1900Myr. In response the orbits widen. The relative change in the inner orbit is larger than that of the outer orbit, such that the ratio $a_{\rm out}/a_{\rm in}$ reduces.  The octupole parameter increases slightly (in absolute terms) and the LK timescale shortens. At the same time (additional) apsidal motion from general relativity, tidal bulges and rotation is reduced. As these terms suppress the LK cycles, reducing the additional apsidal motion allows for 
stronger LK cycles.  With this new configuration, the higher inner eccentricities can be reached that lead to the mass transfer event. In the case of Fig.\,\ref{fig:mt2_ex_wss_hh}, the mass transfer is preceded by orbital “flips”, from retrograde to prograde motion and vice versa \citep{Lit11, Kat11, Tey13,Li14}.	

The evolution in Fig.\,\ref{fig:mt2_ex_wss_hh} is reminiscent of, but alternative to the mass-loss induced eccentric Kozai \citep[MIEK,][]{Sha13, Mic14}. In the MIEK-picture, a triple goes from the quadrupole regime to the octupole regime, and the transfer is a consequence of the change in masses of the stellar components. Typically our systems are in the octupole regime from birth onwards (see also Fig.\,\ref{fig:theory_data}), but the effect of the octupole term is amplified (or less repressed) due to mass loss and its effect on the orbits. 

 Classical MIEK-behaviour where the triples move from the quadrupole to the octupole regime is uncommon in our simulations.  In all models combined, we have only encountered four such systems. The prototype of classical MIEK-behaviour \citep{Sha13, Mic14} is a triple with initially roughly equal mass stellar components, but a large inner mass ratio after the one of the inner stars has become a WD, such that the magnitude of the octopule parameter considerably increases. This phenomenon is uncommon due to the prevalence of low-mass stars in the IMF that do not lose much mass in their winds. We expect MIEK-behaviour to be more likely for more massive stars. Note though that very massive stars  can undergo a supernova explosion with accompanying natal kick, which also drastically changes the dynamical regime of the triple, often leading to a disruption of the system \citep{Pij12, Lu19}.

\begin{figure}
    \centering
    \begin{tabular}{c}
	\includegraphics[clip=true, trim =0mm 15.5mm 0mm 0mm,width=\columnwidth]{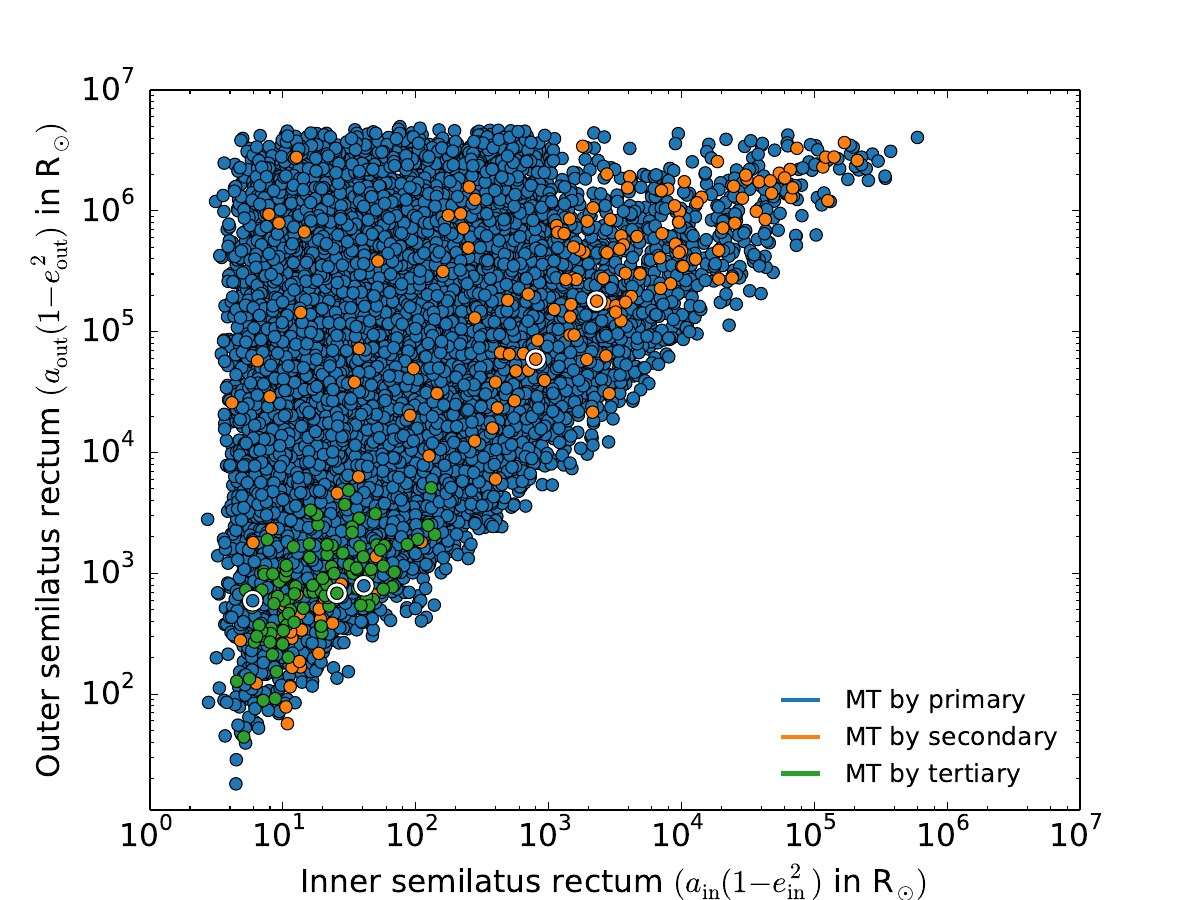} \\[-1mm]
	\includegraphics[clip=true, trim =0mm 0mm 0mm 15.4mm,width=\columnwidth]{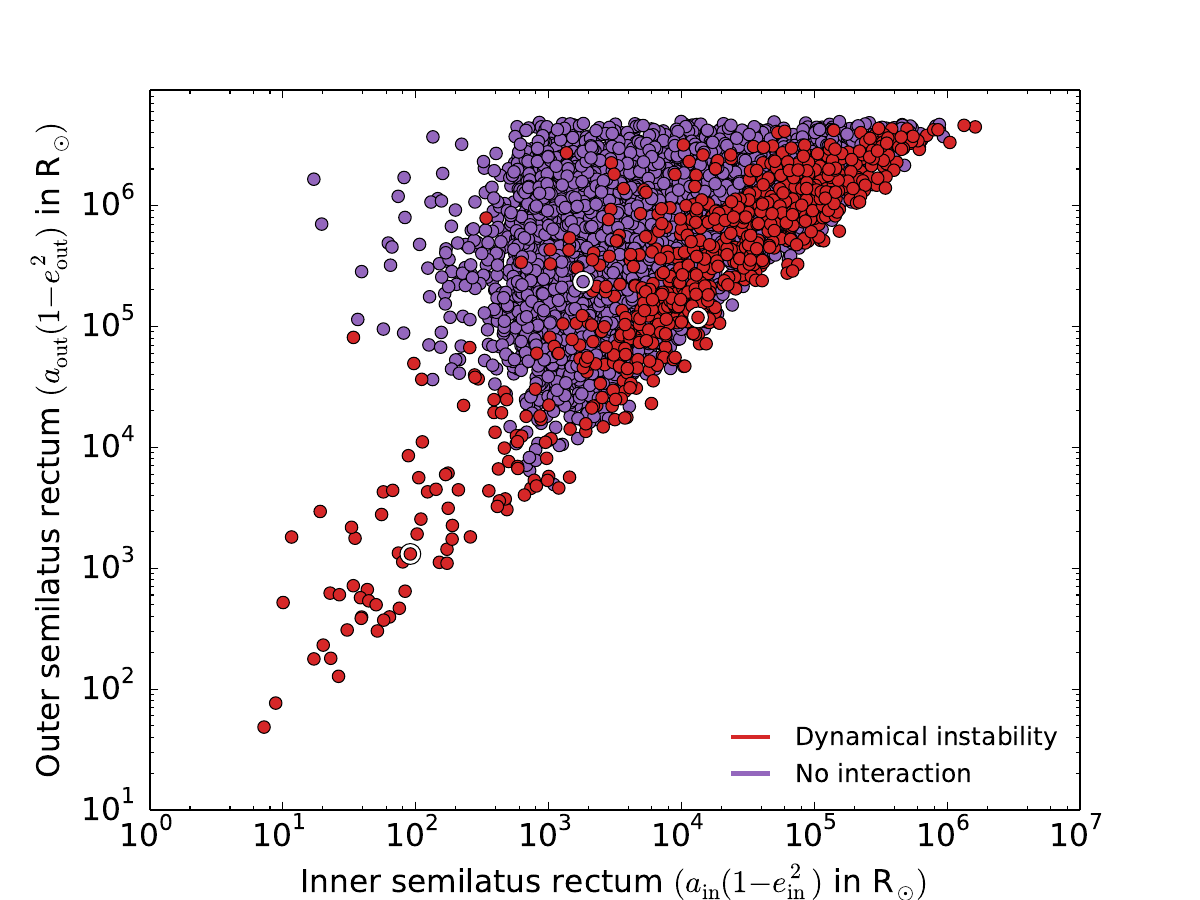} \\
	\end{tabular}
    \caption{Most common evolutionary channels occupy different areas of initial phase space. The x- and and y-axes show the distance the inner and outer orbit would circularise to if tides were efficient. The examplary systems mentioned in the text are marked with white halos.} 
    \label{fig:init_reduced_a}
    \end{figure}

The systems of the second scenario occupy a specific region of phase space on the ZAMS, that is the upper right corner of Fig.\,\ref{fig:init_reduced_a}. As the evolution involves stars that evolve to or through the giant phases, the inner orbits are relatively wide ($a_{\rm in}(1-e_{\rm in}^2) \gtrsim {\rm few} 100\Ro$). The outer orbits are not extremely close to the inner orbit (to ensure dynamical stability), nor are they extremely wide (to ensure that three-body dynamics are relevant on the evolutionary timescale of the stars). The systems of the first scenario (with components on the MS at the onset of mass transfer)  can be found predominantly in the bottom left corner of Fig.\,\ref{fig:init_reduced_a} with ($a_{\rm out}(1-e_{\rm out}^2) \lesssim  10^3-10^4\Ro$). 
 
 \begin{figure}
    \centering
    \begin{tabular}{c}
	\includegraphics[width=\columnwidth]{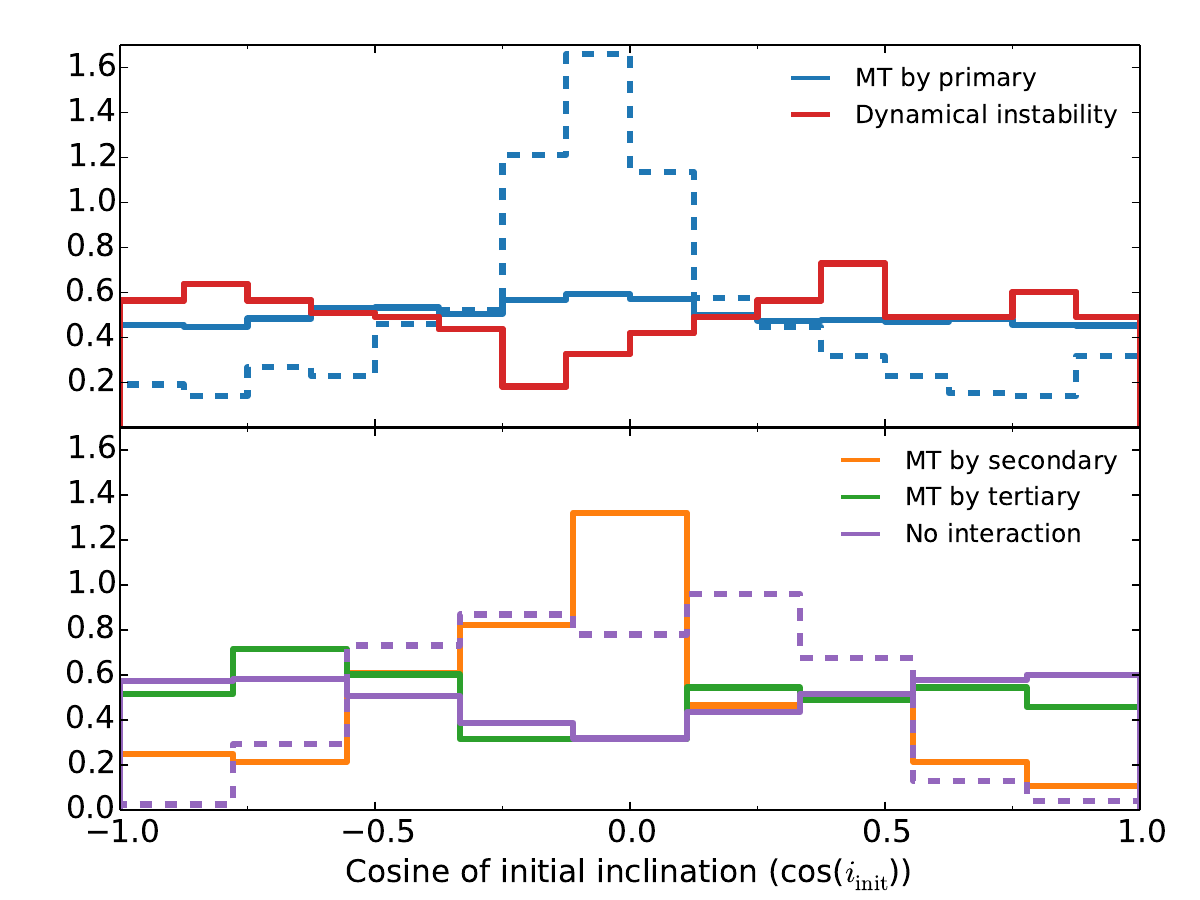} \\
	\end{tabular}
    \caption{Initial distribution of inclinations of the different channels normalised to unity. 
    Systems in which the primary initiates a mass transfer phase (blue) show no stronge bias towards a certain inclination, however, the subset with wide initial orbits ($a_{\rm in}>5000\Ro$) that would not experience mass transfer without the presence of the tertiary are biased towards high initial inclinations. 
Systems in which the secondary initiates a mass transfer phase (orange) are preferentially found at high inclination, whereas systems in which the tertiary initiates mass transfer (green) or there is no mass transfer (purple) preferentially avoid high inclinations. The subsection of the latter that experiences LKCTF are shown as the dashed purple line. The figure is based on model OBin. Model T14 and E09 have similar distributions. 
    } 
    \label{fig:init_incl}
    \end{figure}

 \subsection{Mass transfer initiated by the tertiary star}
  \label{sec:mt3}  
The systems in this channel occupy a specific region of phase space (Fig.\,\ref{fig:theory_data}~and~\ref{fig:init_reduced_a}). Both the inner and the outer orbits are small, such that the tertiary star can fill its Roche lobe. The ratios of the inner and outer semimajor axis are smaller than for those systems where the primary fills its Roche lobe. While this could lead to strong three-body effect, 
LK cycles are not prominent. This can also be seen in Fig.\,\ref{fig:init_incl} where the systems in this channel avoid inclinations around 90$^{\circ}$. If LK cycles were to be important, it would likely lead to mass transfer in the inner binary before the outer star experiences RLOF\footnote{See for instance the example given in section\,\ref{sec:mt1} with initial masses $m_1=1.63\Mo$ and $m_3=2.06\Mo$. Even though $m_1<m_3$, such that the stellar evolution timescale of the primary is longer than that of the tertiary, the primary is the first star to fill the Roche lobe, not due to its expansion in radius, but due to the reduction in the instantenous Roche lobe.}. 

In our study the tertiary star initiates a mass transfer phase in 0.9-1.1\% of triples with an intermediate mass primary.  The rate decreases to 0.6\%  for model OBin as fewer binaries have short outer orbital periods. Our results are consistent with observational estimates.  \cite{DeV14} estimated that out of 725 triples in the Tokovinin catalogue of multiple stellar systems in the Solar neighbourhood \citep{Tok10}, 6 systems ($\approx 1\%$) would experience mass transfer initiated by the outer star. 

We find that initially the tertiary star is more massive than the primary or the secondary star, such that it evolves on a shorter timescale. Typically the donor stars are evolved stars with the far majority of them on the AGB (74-87\%). This is also in agreement with the findings of \cite{DeV14}. The primary and secondary are typically still on the MS. 

The tertiary star is typically less massive than the primary and secondary star combined. In model OBin and Tok14 this is by construction. In model E09, it is the case for 58\% of the systems in this channel. This fraction is set by the initial mass ratio distribution, as well as by stellar winds that strip the envelopes of the evolved tertiary stars.

For 67\% of triples in model OBin and model T14 the outer orbit has circularised (i.e. $e_{\rm out} < 10^{-3}$) at the onset of mass transfer. For model E09 this fraction is lower (i.e. 46\%), but still significant. The systems that maintain eccentric outer orbits upon mass transfer, tend to have high eccentricities with median values between 0.8-0.9. We note that regarding the tides raised on the tertiary star, we assume the inner binary is a point particle. If the tidal perturbation of the tertiary by the (extended) inner binary is efficient (i.e. tertiary tides), it may lead to a contraction of the inner binary and subsequent RLOF or mergers before the tertiary fills the Roche lobe \citep{Gao18}.

A representative example from this channel is the following system: 
$m_1=1.92\Mo$, $m_2=1.47\Mo$, $m_3=2.09\Mo$,   $i=155^{\circ}$,  $a_{\rm in}=28.1\Ro$,  $e_{\rm in}=0.30$, $g_{\rm in} = -0.39 $rad, $a_{\rm out} =1507\Ro$, $e_{\rm out}= 0.74$, $g_{\rm out}= -1.79$rad on the ZAMS, see also Figs.\,\ref{fig:theory_data}~and~\ref{fig:init_reduced_a}.
As the tertiary is more massive than the primary and secondary, it evolves off the MS first. It fills its Roche lobe as an AGB star after 1.32 Gyr. At that time the inner orbit has only changed slightly ($a_{\rm in}=27.6\Ro$,  $e_{\rm in}=0.29$), however the outer orbit has become circularised.  In the absense of stellar winds the outer semi-major axis would have been $a_{\rm in}(1-e_{\rm in}^2) = 681\Ro$, however due to 1.3\% mass loss from the tertiary the final orbit is $a_{\rm out} =690\Ro$.

Next, we explore the outlook of the mass transfer phase and consider if the mass transfer phase is dynamically stable or unstable.
Whereas stable mass transfer in binaries persists on long timescales (e.g. the thermal or nuclear timescale of the donor star), unstable mass transfer occurs on short (dynamical) timescales (i.e. common envelope (CE) evolution), and is expected to lead to orbital contraction and often a merger of the stars. 
 To quantify the stability, we construct the following simple picture. We replace the inner binary with a point particle of the same total mass, we assume the tertiary star has deep surface convection zones, and we assume that the mass transfer is conservative. With these assumptions unstable mass transfer can be avoided if the mass ratio is below a critical mass ratio for which we adopt Eq.57 of \cite{Hur02}. 
Where isolated binaries with evolved donor stars are typically expected to undergo a CE-phase \citep[see e.g.][]{Too14}, 
in this triple channel a common-envelope phase can be avoided for 29-43\% of systems. We stress that our quantification of the stability is a crude model that does not take into account the eccentricity of the orbit, but it suggests that stable mass transfer is more readily achieved in these systems, because the tertiary star is typically less massive than the inner binary (while more massive than the other stars individually).

\begin{figure}
    \centering
	\includegraphics[width=\columnwidth]{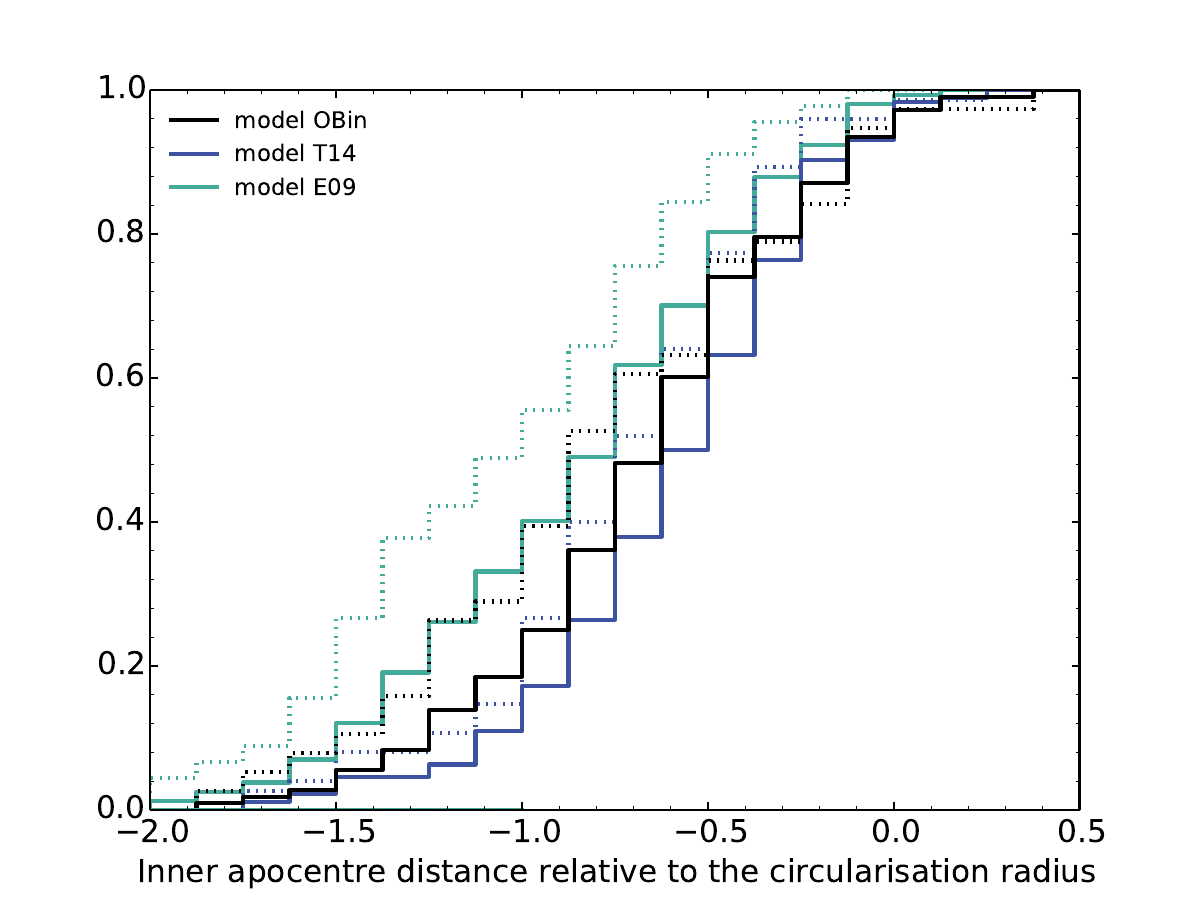} 
    \caption{Investigation into the possibility of the formation of an accretion disc around the inner binary when the tertiary fills its Roche lobe. The figure shows the ratios of the apocentre distance of the inner two stars to the expected circularisation radius of the inflowing material, that is $a_{\rm in}(1+e_{\rm in})/R_{\rm circ}$. The dashed lines represent the distribution for the subset of systems for which we expect the mass transfer from the tertiary towards the inner binary to occur in a stable manner. The distribution is normalised to unity.     }
    \label{fig:rcirc}
    \end{figure}

Subsequently we consider what happens to the material if it flows from the tertiary through the first Lagrangian point towards the inner binary. The material may hit the inner binary directly or  the stream may intersect itself and form an accretion disc around the inner binary. In our systems the classical circularisation radius \citep[See section 4.5 and Eq.4.20 in][]{Fra02} of the material is typically 10-20\% of the outer orbit. When taking into account that an eccentric orbit has less angular momentum by a factor $\sqrt{(1-e_{\rm out}^2)}$, the expected circularisation radius $R_{\rm circ}$ would be smaller by a factor of a few for those orbits. Typically, we find that the inner binaries are of comparable size to the circularisation radius.  We find that only for 11-33\% of systems the inner apocentre is well within the circularisation radius ( $a_{\rm a, in}\equiv a_{\rm in}(1+e_{\rm in})<0.1R_{\rm circ}$),  such that an accretion disc would form (Fig.\,\ref{fig:rcirc}). 
If the proximity of the tertiary star induces or amplifies  the LK cycles and inner eccentricity variations, the fraction of system with accretion disc reduces further. 

Lastly, we address how the stream of material affects the inner binary.
The onset of mass transfer from a tertiary to an inner binary has been simulated by \cite{DeV14} for two observed systems, Tau and HD 97131 using a
combination of stellar evolution, gravitational dynamics and hydrodynamics. In their simulations, \cite{DeV14} finds that an accretion disc is not formed, consistent with our expectations of the triple population. The accretion stream intersect with the orbit of the inner binary and interacts with the stellar companions. A small amount of material is accreted by the stars of the binary, but the majority is ejected from the inner binary. The inner orbital separation shrinks and may ultimately lead to mass transfer in the inner binary itself. The evolution of the mass transfer phase and the dynamical consequences are outside of the scope of the current study \citep[but see ][in the context of twin blue stragglers]{Por19}.

\subsection{Dynamical instability}

One of the possible outcomes is a dynamical instability induced by the tertiary, that is triple evolution dynamical instability (TEDI); initially the triple is dynamically stable, but as the triple evolves it crosses the stability limit (Eq.\,\ref{eq:stab_crit}). The subsequent evolution is reminiscent of that of interactions between a binary and a single star in stellar clusters. It can give rise to close encounters between the stars, a  collision, a compact binary, or a dissolution of the triple in to a binary+single star or three single stars for example.  
This process was also used to explain the great eruption of Eta Carinae \citep{Por16}.

The systems that go through a TEDI are found in a specific region of initial phase space (Fig.\,\ref{fig:init_reduced_a}). 
The initial ratio of outer to inner semilatus rectrum is small, such that the systems are already close to the stability limit at birth. As a result, they are typically in the eccentric LK regime (Fig.\,\ref{fig:theory_data}), with $|\epsilon_{\rm oct}|  > 0.01$ for 75-80\% of TEDI triples and $|\epsilon_{\rm oct}|  > 0.001$ for over 97\%. However, the initial inclinations of the systems are biased away from inclinations around 90$^{\circ}$ to avoid mass transfer in the inner binary (Fig.\,\ref{fig:init_incl}), similar to the previous channel where the tertiary fills its Roche lobe. 

The TEDI occurs in 2.3-2.5\% of all triples in model T14 and E09, and 4.2\% in model OBin. The birthrate in model OBin is increased as there are more systems with wide inner orbits initially  ($a_{\rm in}\gtrsim 10^3\Ro$).

We differentiate between two scenarios. In the first scenario the dynamical instability occurs when all stars are still on the MS. The system goes through many LK cycles (of the order of tens to thousands of cycles on average). Due to the higher-order terms in the secular approximation, the outer eccentricity increases until the system crosses the stability limit. In Fig.\,\ref{fig:theory} the system would move horizontally towards higher outer eccentricities. 
In the second scenario, the primary star has moved of the MS. During the giant stages of evolution, the primary star suffers strong mass losses through stellar winds. Assuming an adiabatic effect on the orbit, the orbit widens in response to the mass loss. The relative amount of orbital widening is proportional to the relative amount of mass loss. This means that the inner orbit widens more relatively, approaches the outer orbit, and the system as a whole can become unstable \citep{Kis94, Ibe99, Per12}.
In Fig.\,\ref{fig:theory} the system would move vertically downwards.
At the moment of the TEDI, the primary is typically an evolved star. For 62-64\% of systems the primary is on the AGB. In 24-31\% of systems, the primary has already become a WD, and the systems becomes unstable due to the stellar winds that acted on the orbit as well as enhanced the importance of the octupole terms.

The systems in the second scenario have initial orbital separations of $a_{\rm in} (1-e_{\rm in}^2) \gtrsim 10^3 \Ro$ (Fig.\,\ref{fig:init_reduced_a}).  Systems evolving through the first scenario have initial semilatus recti that extend to smaller values. An example of this evolution is marked in Figs.\,\ref{fig:theory_data}~and~\ref{fig:init_reduced_a}. The initial conditions are $m_1 = 1.25\Mo$, $m_2 = 0.52\Mo$, $m_3=0.57\Mo$, $i=75.4^{\circ}$, $a_{\rm in} =291\Ro$, $e_{\rm in}=0.83$,  $g_{\rm in}= -0.93$rad, $a_{\rm out} = 1427\Ro$, $e_{\rm out} = 0.29$, $g_{\rm out}= 1.73$rad. At birth the system is close to the stability limit, and crosses it after 181Myr. On this timescale the stars do not loose any mass to the stellar winds. With an LK timescale of 379yr, the system has undergone thousands of LK cycles before crossing the stability boundary. The system becomes dynamically unstable as the outer eccentricity increases from the initial 0.29 to 0.34 due to interactions with the tertiary star (indicated by the arrow in Fig.\,\ref{fig:theory_data}). The system is well inside the octupole regime with an octupole parameter of 0.026-0.031.

Figs.\,\ref{fig:theory_data}~and~\ref{fig:init_reduced_a} also show an example of the second scenario (with initial parameters $m_1 = 2.72\Mo$, $m_2 = 0.59\Mo$, $m_3=1.29\Mo$, $i=6.2^{\circ}$, $a_{\rm in} =13760\Ro$, $e_{\rm in}=0.18$,  $g_{\rm in}= 0.68$rad, $a_{\rm out} = 174881\Ro$, $e_{\rm out} = 0.57$, $g_{\rm out}= -0.51$rad).
As the primary star evolves and ascends the AGB, it loses mass to stellar winds. After 632 Myr, the primary has lost 45\% of its mass and the system crosses the stability limit. Due to the wind mass loss, the inner orbit has widened more than the outer orbit with factors 1.58 and 1.36 respectively. The system moves vertically downwards in Fig.\,\ref{fig:theory_data} by  14\%. This is within the size of the marker.

The fraction of systems with unevolved primaries is highest in model T14 and E09 (56\% and 50\% respectively) compared to model OBin (37\%), as there are more orbits that are compact initially in model T14 and E09.
When the system becomes unstable, the secondary and tertiary are typically still on the MS in both scenarios.

We find an galactic TEDI rate of 1-2 per 1000 years, which is in agreement with the estimates of \cite{Per12} (4-6 per 1000 years) and \cite{Ham13} (0.4-0.75 per 1000 years). 
The former is based on a hybrid method combining analytical estimates with triple population synthesis excluding LK effects, and the latter is based on a population synthesis study of wide triples. 
Furthermore, \cite{Per12} simulate the further evolution of the triple, and show that for roughly a third of the systems destabilise on a short timescale (0.5Myr), and subsequently 10\% of systems experience a collision between two of the stars of the triple. The stellar collision rate from destabilising isolated triples is an order of magnitude higher than that due to random encounters in Galactic globular clusters.

\subsection{Survival for a Hubble time}
In this evolutionary channel the systems stay intact for a Hubble time without experiencing RLOF and without becoming dynamically unstable. In the mass range considered in this study, the primary stars become WDs within a Hubble time. This channel is the second most common pathway for triples. About 27-30\% of triples evolve this way in model OBin and E09, and 20\% for model T14. The channel is less likely in model T14, as the average orbit tend to be more compact compared to the other models, making it harder to avoid RLOF (see also Fig.\,\ref{fig:init_a}).

The systems of the current channel  have wide inner and outer orbits (Fig.\,\ref{fig:init_reduced_a}). Initially the outer orbits $a_{\rm out}(1-e_{\rm out}^2) \gtrsim 10^5\Ro$. The median size of the outer orbit varies between the models, as they are closely linked to the initial distribution of orbits (see also Fig.\,\ref{fig:init_a}). Initially the inner orbits range from $a_{\rm in}(1-e_{\rm in}^2) \approx 500-10^5\Ro$ and peak around $2000-4000\Ro$. RLOF is avoided by definition in this channel. 

A representative example of a system in this channel is marked in Figs.\,\ref{fig:theory_data}~and~\ref{fig:init_reduced_a}. On the ZAMS it has the following parameters: 
$m_1=1.60\Mo$, $m_2=0.75\Mo$, $m_3=1.07\Mo$,   $i=16^{\circ}$,  $a_{\rm in}=2472\Ro$,  $e_{\rm in}=0.51$, $g_{\rm in} = 0.25 $rad, $a_{\rm out} =356776\Ro$, $e_{\rm out}= 0.59$, $g_{\rm out}= 1.55$rad. In this system, the primary evolves off the MS first, sheds its envelope during the giant phases, and becomes a WD around $2.5 Gyr$. It this point the inner and outer orbits have widened to $4058\Ro$ and $506771\Ro$, respectively. Subsequently the tertiary evolves and becomes a WD around $9.8 Gyr$. As we assume the mass lost from the tertiary does not interact with the inner orbit, only the outer orbit is affected, and it widens to  $654756 \Ro$. The secondary star is of sufficiently low-mass to not evolve off the MS in a Hubble time. And thus, at the end of the simulation this sytem consists of a wide WD-MS binary which is  orbited by a second WD. 

Observational counterparts of this channel are the WD systems WD0413-07 (40 Eridani) and WD2351--335 \citep{Weg88, Sch04, Far05, Tok08, Too17}. 40 Eridani B is one of the closest WDs to us \citep[e.g.][]{Hol18}. It has a M-dwarf companion in an orbit of 11.9" and a K-star companion in a wider orbit of 83.4".  WD2351-335 has two M-dwarf companions at 6.6" and 103" respectively. 

Such systems with a single WD in the inner binary are relatively common in our simulations; 20-30\% of systems in this channel. In 7-11\% the secondary star becomes a WD as well and a wide double white dwarf binary is formed.  For 30-41\% of systems only the primary and the tertiary star becomes WDs. This option is more common in model E09, as more tertiaries are initially massive enough  to form a WD within a Hubble time (Fig.\,\ref{fig:init_q}). Lastly, 29-33\% of systems in all models from a wide triple WD, as recently observed for the first time by \cite{Per19}.

As the stellar components evolve from the ZAMS to the remnant stages, their orbits change due to mass loss from stellar winds. Assuming mass loss occurs adiabatically, the orbits widen. In our simulations the outer orbit widens within a factor $\sim$5 typically. Many inner orbits widen as well, however a reduction of the inner orbit is also possible due to LKCTF. In 11-13\% of systems in this channel the semilatus rectum of the inner orbit has decreased after a Hubble time, and 6-7\%  reach complete circularisation. 

The initial distribution of inclinations is shown in Fig\,\ref{fig:init_incl}. Initial inclinations around 90$^{\circ}$ are avoided, as strong LK cycles would lead to RLOF. The systems that experience LKCTF-like behaviour and inner orbital shrinkage, are preferentially born at initial inclinations closer to 90$^{\circ}$ (purple dashed line in Fig.\,\ref{fig:init_incl}). 
Their inclinations oscillate and typically freeze-out around $40-50^{\circ}$ and $130-140^{\circ}$ \citep[see also ][for low-mass stellar triples]{Fab07, Nao14}

\section{Discussion} 
\label{sec:discussion}
In this study we have stellar triples by modelling their evolution in a comprehensive way. That is we take into account both stellar evolution as well as three-body dynamics, and include the effects of tides and stellar wind mass losses. 
Here, we critically discuss the main assumptions that we did not vary in out modelling, and their effects on triple evolution.

\textit{Distribution of initial inclinations:} In this study we have assumed an uniform distribution in the cosine of the inclination. Observationally, this is a good approximation for outer projected separations\footnote{ We note that on average the difference between the projected separation and the semi-major axis of the orbit is small. Assuming an average sky inclination of 60$^{\circ}$ (from a random distribution on the sky), and an average eccentricity of $2/3$ (from the thermal distribution), the projected separation is 78\% of the semi-major axis.}
that are $\gtrsim 10^5\Ro$. However, at $\lesssim 10^4\Ro$  orbits are more aligned \citep{Tok17}. This represents about 10\% of systems in model OBin and E09, and about 20\% in T14. It corresponds to the evolutionary channels where either the primary or the tertairy fills its Roche lobe (Fig.\,\ref{fig:init_reduced_a}), where the latter preferably occurs in less inclined systems. Therefore, if we would incorporate the orbital alignment in triples with compact outer orbits, we expect that tertiary will be able to fill its Roche lobe in an even larger fraction of triples.

\textit{Distribution of initial eccentricities:} In our models we have assumed that the initial eccentricities are distributed thermally ($f(e) \propto 2e$). There are observational indications that the distribution is flatter \cite[e.g.][]{Tok16,Moe17}. In the absense of three-body dynamics, the effect would be similar to that of taking a different distribution of orbital separations that is biased towards wider orbits. This would reduce the number of systems in which the primary or tertiary fills its Roche lobe (Fig.\,\ref{fig:init_reduced_a}). Regarding three-body dynamics, we expect some differences as well. If the outer eccentricities are smaller on average, one would move to the left in Fig.\,\ref{fig:theory}. The minimum 
$a_{\rm out}/a_{\rm in}$ while preserving dynamical stability decreases. This would enhance the number of systems where the tertiary fills the Roche lobe. Also the parameter space of the standard LK regime is larger, and the LK-timescales themselves are longer as well, which means there would be more triples that do not interact and stay intact for a Hubble time. 
The LK cycles would  be effected further through their dependence on the initial  inner eccentricity. 
For initially circularised inner orbits in the test-particle approximation, LK cycles can only take place when the inclination is between 39.2-140.8$^{\circ}$ with a maximum eccentricity given by Eq.\,\ref{eq:e_max}. For non-zero eccentricities, the parameter space widens in which LK cycles take place, and higher maximum eccentricties can be reached (depending on the initial eccentricity and the argument of pericentre).  If the initial inner eccentricities are smaller on average, the occurence of LK cycles and their amplitudes is likely reduced.

\textit{Distribution of initial mass ratios:}
In model OBin \& T14, we have adopted a simple formalism for the mass ratio distribution based on a first-order approximation to observations. This formalism consists of a uniform mass ratio distribution for the inner and outer mass ratios. The modelling of \citep{Egg09} includes a populations of 'twins' , that is near-equal-mass pairs, for compact orbits ($P<25$d). They make up 25\% of the population of compact binaries. This does not strongly affect how many triples experience mass transfer, but it does affect the subsequent mass transfer phase (for example the stability of the mass transfer and the orbital evolution). 

There are also indications that the mass ratio distributions in wide binaries deviate from the uniform mass ratio distributions that are observed in more compact systems \citep[e.g.][]{Moe17}. The distributions for wide binaries would include more smaller companion masses than given by a uniform mass ratio distribution. A detailed study on the effect of this for triple evolution is beyond the scope of the current study, but our expectations are the following: First, fewer tertiaries would evolve off the main-sequence in a Hubble time. Therefore, naively, one would expect also fewer tertiaries to fill their Roche lobe. However, the systems in which the tertiary fills its Roche lobe have compact orbits initially, and therefore the occurrence rate of mass transfer that is initiated by the tertiary are likely well approximated in this work. 
Secondly, if the tertiary mass decreases (and other parameters stay the same), the LK timescales would increase, and the dynamical impact of the tertairy on the inner binary would decrease. 
In a previous study using the same code \texttt{TRES} \citep{Too18} which focused on wide triples ($a_{\rm in}(1-e_{\rm in}^2) > 2500 R_{\odot}$) in a similar mass range ($m_1 , m_2, m_3 < 7.7M_{\odot}$; $m_1, m_2 > 0.95M_{\odot}$), we varied the mass ratio distributions of the inner and outer orbits. 
Under the extreme assumption that the tertiary mass in uncorrelated to the mass of the inner two stars (i.e. drawn from a Kroupa-IMF), the fraction of systems undergoing mass transfer slightly decreased from 10-12$\%$  (their model  Q\_OUT ) to 9.3-9.9$\%$  (their model STD).

\textit{Tidal evolution:} 
Our simulations are based on a single set of assumptions for tidal interactions and their timescales, which affect 
 the average eccentricities upon RLOF and the degree of orbital shrinkage during LKCTF.
We note that our modelling is based on the classical equilibrium-tide model \cite{Hut81}, which may not be appropriate for inner orbits with large eccentricities ($e_{\rm in}\gtrsim 0.8$) and small periastron distances ($\lesssim 5R_{\star}$). In this regime, tidal energy dissipation occurs solely near periastron, where as in the equilibrium-tide model it is averaged over the entire orbit. Tidal energy dissipations is dominated by non-radial dynamical oscillations in stead, which are more efficient by orders of magnitude \citep[e.g.][]{Pre77, McM86, Koc92, Mar95}. 
The transition between the two regimes occurs near $e \approx 0.8$ \citep{Mar95}. 
The highly eccentric inner orbits of triples are expected to decay to smaller eccentricities ($\approx 0.6-0.8$) at roughly constant periastron distance \citep{Mar95, Moe18}. 
As the instanteneous Roche lobe of a star is roughly proportional to the periastron distance \citep{Egg83, Sep07},
the time of RLOF is likely not affected much. The inner orbit would be somewhat smaller and less eccentric.

\textit{Stellar winds:}
In the current study we have made simplifying assumptions regarding the stellar winds; we assume that stars loose mass in fast, spherically symmetric winds and that the wind matter is not accreted by any of the companion stars.  
The former is a common assumption in binary population synthesis studies \citep[see e.g. the appendix of ][for an overview]{Too14}, the latter is a simplification we have adopted in the current study. It mostly affects systems in the last three channels, as intermediate mass stars mostly suffer from stellar winds in the late stages of their evolution.

If the wind matter forms a more or less uniform medium around the system, one would expect Bondi-Hoyle-Littleton accretion \citep{Bon44} to take place. The amount of accretion on a stellar companion from the wind depends on the orbital parameters, the wind mass rate and wind velocity of the mass-losing star. By ignoring the possibility of wind accretion in this work, our simulations underestimate the companion masses up a few percents and overestimate the orbital separations accordingly. If the wind and orbital velocities are comparable in magnitude, the assumption of spherical symmetry is violated. In this case the wind can be focused and accreted on to the companion star with an enhanced efficiency and a possible contraction of the orbit \citep[e.g.][]{Moh07, Sal18, Che20}. Furthermore, \cite{Com19} showed that the efficiency of wind accretion from a tertiary star onto the stars in the inner binary may be enhanced up to a factor two compared to the case where the secondary star was removed. 
We defer a more detailed investigation of the potential effects of wind accretion to a future paper.

\textit{Numerical stability:}
We note that in a few cases our numerical solver was unable to find a solution for the dynamical evolution of the systems.  It concerns 0.35\% of systems in model OBin, 0.65\% for model T14, and 0.13\% for model E09. 
The systems are compact  with outer semilatus rectum smaller than a few thousand solar radii (bottom left in Fig.\,\ref{fig:init_reduced_a}). The primary star is close to filling its Roche lobe, such that the tidal torques are strong.  The systems have short LK timescales (i.e. $\lesssim 10^4$yr) compared to a Hubble time or the stellar lifetimes, and as such will undergo numerous LK cycles.  Higher-order terms of the secular approximation are important for their evolution as well, that is $\epsilon_{\rm oct} \gtrsim 0.001-0.01$.

\section{Conclusions} 
\label{sec:concl}

In this paper we aim to provide a systematic exploration of the evolution of stellar triples. We focus on the long-term evolution of isolated and hierarchical triples with low- and intermediate mass stars. The evolution of these systems are governed by stellar evolution, three-body dynamics, and dissipative processes such as tides and stellar winds. For a review of the relevant processes see \cite{Too16}. The expected birth rate of these triples is $\sim5$ per decade, and  $\sim50$ per decade in the full mass range.  

We find that in the majority of triples, the stellar components interact with each other during their lifetime (Tbl.\,\ref{tbl:birthrates}, Fig.\,\ref{fig:pie}). This can be either dynamically (through disruptions of the systems, or modifications to the stellar orbits) or physically (through mass transfer, mergers or collisions). 
Typically the initially most massive star in the inner binary fills its Roche lobe and starts transferring mass to its companion(s).
This happens in 63-74\% of triples in our samples depending on the model. A few percent of triples (2-4.5\%) becomes dynamically unstable during their evolution. In less then a percent of triples the secondary or tertiary fills its Roche lobe before any other star does. One interesting sub-group is that were the secondary star starts transferring mass to a compact object. With a Galactic birthrate of 1-2 per 10000yrs, these sources can give rise to novae and x-ray emission without any prior phase of mass transfer or common-envelope evolution. 

When we compare our evolving triple population with a population of pure binaries in the same mass range, we find that mass transfer takes place in a larger percentage of triples; 66-77\% versus 28-39\% respectively. This is in part because the presence of the tertiary star drives the inner binary to mass transfer, but mostly because the inner binaries of triples tend to be more compact compared to isolated binaries (Fig.\,\ref{fig:init_a}). 
Another difference between triple and binary evolution is that in triples the orbits are not typically circularised upon mass transfer. About 40\% of cases the orbit is still significantly eccentric (Figs.\,\ref{fig:mt1_ecc_hist}~and~\ref{fig:mt1_donor_hist}). Typically when the donor stars are fairly unevolved  upon Roche lobe overflow (e.g. on the main-sequence), the orbits are eccentric, and vice versa.

In this paper, we focus on triple evolution up onto the first phase of mass transfer, but if the triple system survives the interaction, more mass transfer phases can follow. We leave the outcome of the mass transfer phases to forthcoming papers. Assuming that mass transfer often leads to a stellar merger or dissolution of the triple only about 20-30\% of triples survive for a Hubble time.

Our main conclusions are relatively robust to uncertainties in the primordial population of triples (in contrast with binary evolution). Eventhough we start with very different input distributions for the orbits in the triples, the resulting initial distributions are more similar due to the removal of dynamical unstable systems. Such systems typically dissolve on dynamical timescales such that stellar evolution is less relevant.   

We have also provided a simple tool (Fig.\,\ref{fig:theory}) with which one can estimate the dynamical impact of the tertiary star on the inner binary - for any given triple. An interactive version of Fig.\,\ref{fig:theory}a is available at \href{https://bndr.it/wr64f}{\color{blue}https://bndr.it/wr64f}. 
The graphical interface allows the user to adjust the figure to any given triple.

\begin{acknowledgements}
ST acknowledges support from the Netherlands Research Council NWO (grant VENI [nr. 639.041.645]). 
A.S.H. gratefully acknowledges support from the Institute for Advanced Study, the Peter Svennilson Membership, and the Martin A. and Helen Chooljian Membership. 
\end{acknowledgements}

\bibliographystyle{aa}
\bibliography{bibtex_silvia_toonen}

\hyphenation{Post-Script Sprin-ger}
\begin{thebibliography}{146}
\expandafter\ifx\csname natexlab\endcsname\relax\def\natexlab#1{#1}\fi

\bibitem[{{Abt}(1983)}]{Abt83}
{Abt}, H.~A. 1983, \araa, 21, 343

\bibitem[{{Antognini} {et~al.}(2014){Antognini}, {Shappee}, {Thompson}, \&
  {Amaro-Seoane}}]{Ant14b}
{Antognini}, J.~M., {Shappee}, B.~J., {Thompson}, T.~A., \& {Amaro-Seoane}, P.
  2014, \mnras, 439, 1079

\bibitem[{{Antognini}(2015)}]{Ant15}
{Antognini}, J.~M.~O. 2015, \mnras, 452, 3610

\bibitem[{{Antonini} {et~al.}(2014){Antonini}, {Murray}, \& {Mikkola}}]{Ant14}
{Antonini}, F., {Murray}, N., \& {Mikkola}, S. 2014, \apj, 781, 45

\bibitem[{{Antonini} \& {Perets}(2012)}]{Ant12}
{Antonini}, F. \& {Perets}, H.~B. 2012, \apj, 757, 27

\bibitem[{{Antonini} {et~al.}(2017){Antonini}, {Toonen}, \& {Hamers}}]{Ant17}
{Antonini}, F., {Toonen}, S., \& {Hamers}, A.~S. 2017, \apj, 841, 77

\bibitem[{{Bataille} {et~al.}(2018){Bataille}, {Libert}, \& {Correia}}]{Bat18}
{Bataille}, M., {Libert}, A.-S., \& {Correia}, A.~C.~M. 2018, \mnras, 479, 4749

\bibitem[{{Belczynski} {et~al.}(2002){Belczynski}, {Kalogera}, \&
  {Bulik}}]{Bel02}
{Belczynski}, K., {Kalogera}, V., \& {Bulik}, T. 2002, \apj, 572, 407

\bibitem[{{Blaes} {et~al.}(2002){Blaes}, {Lee}, \& {Socrates}}]{Bla02}
{Blaes}, O., {Lee}, M.~H., \& {Socrates}, A. 2002, \apj, 578, 775

\bibitem[{{Bobrick} {et~al.}(2017){Bobrick}, {Davies}, \& {Church}}]{Bob17}
{Bobrick}, A., {Davies}, M.~B., \& {Church}, R.~P. 2017, \mnras, 467, 3556

\bibitem[{{Bode} \& {Wegg}(2014)}]{Bod14}
{Bode}, J.~N. \& {Wegg}, C. 2014, \mnras, 438, 573

\bibitem[{{Boffin} {et~al.}(2014){Boffin}, {Hillen}, {Berger}, {Jorissen},
  {Blind}, {Le Bouquin}, {Miko{\l}ajewska}, \& {Lazareff}}]{Bof14}
{Boffin}, H.~M.~J., {Hillen}, M., {Berger}, J.~P., {et~al.} 2014, \aap, 564, A1

\bibitem[{{Bona{\v c}i{\'c} Marinovi{\'c}} {et~al.}(2008){Bona{\v c}i{\'c}
  Marinovi{\'c}}, {Glebbeek}, \& {Pols}}]{Bon08}
{Bona{\v c}i{\'c} Marinovi{\'c}}, A.~A., {Glebbeek}, E., \& {Pols}, O.~R. 2008,
  \aap, 480, 797

\bibitem[{{Bondi} \& {Hoyle}(1944)}]{Bon44}
{Bondi}, H. \& {Hoyle}, F. 1944, \mnras, 104, 273

\bibitem[{{Borkovits} {et~al.}(2010){Borkovits}, {Csizmadia}, {Paragi},
  {Sturmann}, {Sturmann}, {Farrington}, {McAlister}, {ten Brummelaar}, \&
  {Turner}}]{Bor10}
{Borkovits}, T., {Csizmadia}, S., {Paragi}, Z., {et~al.} 2010, in Astronomical
  Society of the Pacific Conference Series, Vol. 435, Binaries - Key to
  Comprehension of the Universe, ed. A.~{Pr{\v s}a} \& M.~{Zejda}, 217

\bibitem[{{Borkovits} {et~al.}(2004){Borkovits}, {Forg{\'a}cs-Dajka}, \&
  {Reg{\'a}ly}}]{Bor04}
{Borkovits}, T., {Forg{\'a}cs-Dajka}, E., \& {Reg{\'a}ly}, Z. 2004, \aap, 426,
  951

\bibitem[{{Chen} {et~al.}(2020){Chen}, {Ivanova}, \&
  {Carroll-Nellenback}}]{Che20}
{Chen}, Z., {Ivanova}, N., \& {Carroll-Nellenback}, J. 2020, \apj, 892, 110

\bibitem[{{Claeys} {et~al.}(2014){Claeys}, {Pols}, {Izzard}, {Vink}, \&
  {Verbunt}}]{Cla14}
{Claeys}, J.~S.~W., {Pols}, O.~R., {Izzard}, R.~G., {Vink}, J., \& {Verbunt},
  F.~W.~M. 2014, \aap, 563, A83

\bibitem[{{Comerford} {et~al.}(2019){Comerford}, {Izzard}, {Booth}, \&
  {Rosotti}}]{Com19}
{Comerford}, T.~A.~F., {Izzard}, R.~G., {Booth}, R.~A., \& {Rosotti}, G. 2019,
  \mnras, 490, 5196

\bibitem[{{de Kool} \& {Ritter}(1993)}]{deK93}
{de Kool}, M. \& {Ritter}, H. 1993, \aap, 267, 397

\bibitem[{{de Vries} {et~al.}(2014){de Vries}, {Portegies Zwart}, \&
  {Figueira}}]{DeV14}
{de Vries}, N., {Portegies Zwart}, S., \& {Figueira}, J. 2014, \mnras, 438,
  1909

\bibitem[{{Dosopoulou} \& {Kalogera}(2016)}]{Dou16b}
{Dosopoulou}, F. \& {Kalogera}, V. 2016, \apj, 825, 71

\bibitem[{{Duch{\^e}ne} \& {Kraus}(2013)}]{Duc13}
{Duch{\^e}ne}, G. \& {Kraus}, A. 2013, \araa, 51, 269

\bibitem[{{Duquennoy} \& {Mayor}(1991)}]{Duq91}
{Duquennoy}, A. \& {Mayor}, M. 1991, \aap, 248, 485

\bibitem[{{Eggleton}(1983)}]{Egg83}
{Eggleton}, P.~P. 1983, \apj, 268, 368

\bibitem[{{Eggleton}(2009)}]{Egg09}
{Eggleton}, P.~P. 2009, \mnras, 399, 1471

\bibitem[{{Eggleton} \& {Tokovinin}(2008)}]{Egg08}
{Eggleton}, P.~P. \& {Tokovinin}, A.~A. 2008, \mnras, 389, 869

\bibitem[{{Eldridge}(2009)}]{Eld09}
{Eldridge}, J.~J. 2009, \mnras, 400, L20

\bibitem[{{Fabrycky} \& {Tremaine}(2007)}]{Fab07}
{Fabrycky}, D. \& {Tremaine}, S. 2007, \apj, 669, 1298

\bibitem[{{Farihi} {et~al.}(2005){Farihi}, {Becklin}, \& {Zuckerman}}]{Far05}
{Farihi}, J., {Becklin}, E.~E., \& {Zuckerman}, B. 2005, \apjs, 161, 394

\bibitem[{{Ford} {et~al.}(2000){Ford}, {Kozinsky}, \& {Rasio}}]{For00}
{Ford}, E.~B., {Kozinsky}, B., \& {Rasio}, F.~A. 2000, \apj, 535, 385

\bibitem[{{Fragione} \& {Kocsis}(2019)}]{Fra19b}
{Fragione}, G. \& {Kocsis}, B. 2019, \mnras, 486, 4781

\bibitem[{{Fragione} \& {Loeb}(2019)}]{Fra19}
{Fragione}, G. \& {Loeb}, A. 2019, \mnras, 486, 4443

\bibitem[{{Frank} {et~al.}(2002){Frank}, {King}, \& {Raine}}]{Fra02}
{Frank}, J., {King}, A., \& {Raine}, D.~J. 2002, {Accretion Power in
  Astrophysics: Third Edition}, 398

\bibitem[{{Gao} {et~al.}(2018){Gao}, {Correia}, {Eggleton}, \& {Han}}]{Gao18}
{Gao}, Y., {Correia}, A.~C.~M., {Eggleton}, P.~P., \& {Han}, Z. 2018, \mnras,
  479, 3604

\bibitem[{{Gualandris} {et~al.}(2004){Gualandris}, {Portegies Zwart}, \&
  {Eggleton}}]{Gua04}
{Gualandris}, A., {Portegies Zwart}, S., \& {Eggleton}, P.~P. 2004, \mnras,
  350, 615

\bibitem[{{Haim} \& {Katz}(2018)}]{Hai18}
{Haim}, N. \& {Katz}, B. 2018, \mnras, 479, 3155

\bibitem[{{Hamers}(2019)}]{Ham19d}
{Hamers}, A.~S. 2019, \mnras, 482, 2262

\bibitem[{{Hamers} \& {Dosopoulou}(2019)}]{Ham19}
{Hamers}, A.~S. \& {Dosopoulou}, F. 2019, \apj, 872, 119

\bibitem[{{Hamers} {et~al.}(2013){Hamers}, {Pols}, {Claeys}, \&
  {Nelemans}}]{Ham13}
{Hamers}, A.~S., {Pols}, O.~R., {Claeys}, J.~S.~W., \& {Nelemans}, G. 2013,
  \mnras, 430, 2262

\bibitem[{{Hamers} \& {Thompson}(2019{\natexlab{a}})}]{Ham19b}
{Hamers}, A.~S. \& {Thompson}, T.~A. 2019{\natexlab{a}}, \apj, 883, 23

\bibitem[{{Hamers} \& {Thompson}(2019{\natexlab{b}})}]{Ham19c}
{Hamers}, A.~S. \& {Thompson}, T.~A. 2019{\natexlab{b}}, \apj, 882, 24

\bibitem[{{Han} {et~al.}(2002){Han}, {Podsiadlowski}, {Maxted}, {Marsh}, \&
  {Ivanova}}]{Han02}
{Han}, Z., {Podsiadlowski}, P., {Maxted}, P.~F.~L., {Marsh}, T.~R., \&
  {Ivanova}, N. 2002, \mnras, 336, 449

\bibitem[{{Harrington}(1968)}]{Har68}
{Harrington}, R.~S. 1968, \aj, 73, 190

\bibitem[{{Heggie}(1975)}]{Heg75}
{Heggie}, D.~C. 1975, \mnras, 173, 729

\bibitem[{{Hollands} {et~al.}(2018){Hollands}, {Tremblay}, {G{\"a}nsicke},
  {Gentile-Fusillo}, \& {Toonen}}]{Hol18}
{Hollands}, M.~A., {Tremblay}, P.-E., {G{\"a}nsicke}, B.~T., {Gentile-Fusillo},
  N.~P., \& {Toonen}, S. 2018, \mnras, 480, 3942

\bibitem[{{Huang}(1956)}]{Hua56}
{Huang}, S.~S. 1956, \aj, 61, 49

\bibitem[{{Huang}(1963)}]{Hua63}
{Huang}, S.-S. 1963, \apj, 138, 471

\bibitem[{{Hurley} {et~al.}(2000){Hurley}, {Pols}, \& {Tout}}]{Hur00}
{Hurley}, J.~R., {Pols}, O.~R., \& {Tout}, C.~A. 2000, \mnras, 315, 543

\bibitem[{{Hurley} {et~al.}(2002){Hurley}, {Tout}, \& {Pols}}]{Hur02}
{Hurley}, J.~R., {Tout}, C.~A., \& {Pols}, O.~R. 2002, \mnras, 329, 897

\bibitem[{{Hut}(1980)}]{Hut80}
{Hut}, P. 1980, \aap, 92, 167

\bibitem[{{Hut}(1981)}]{Hut81}
{Hut}, P. 1981, \aap, 99, 126

\bibitem[{{Iben} \& {Tutukov}(1999)}]{Ibe99}
{Iben}, Jr., I. \& {Tutukov}, A.~V. 1999, \apj, 511, 324

\bibitem[{{Innanen} {et~al.}(1997){Innanen}, {Zheng}, {Mikkola}, \&
  {Valtonen}}]{Inn97}
{Innanen}, K.~A., {Zheng}, J.~Q., {Mikkola}, S., \& {Valtonen}, M.~J. 1997,
  \aj, 113, 1915

\bibitem[{{Izzard} {et~al.}(2009){Izzard}, {Glebbeek}, {Stancliffe}, \&
  {Pols}}]{Izz09}
{Izzard}, R.~G., {Glebbeek}, E., {Stancliffe}, R.~J., \& {Pols}, O.~R. 2009,
  \aap, 508, 1359

\bibitem[{{Kashi} \& {Soker}(2018)}]{Kas18}
{Kashi}, A. \& {Soker}, N. 2018, \mnras, 480, 3195

\bibitem[{{Katz} \& {Dong}(2012)}]{Kat12}
{Katz}, B. \& {Dong}, S. 2012, ArXiv e-prints

\bibitem[{{Katz} {et~al.}(2011){Katz}, {Dong}, \& {Malhotra}}]{Kat11}
{Katz}, B., {Dong}, S., \& {Malhotra}, R. 2011, Physical Review Letters, 107,
  181101

\bibitem[{{Kervella} {et~al.}(2017){Kervella}, {Th{\'e}venin}, \&
  {Lovis}}]{Ker17}
{Kervella}, P., {Th{\'e}venin}, F., \& {Lovis}, C. 2017, \aap, 598, L7

\bibitem[{{Kinoshita} \& {Nakai}(1999)}]{Kin99}
{Kinoshita}, H. \& {Nakai}, H. 1999, Celestial Mechanics and Dynamical
  Astronomy, 75, 125

\bibitem[{{Kippenhahn} \& {Weigert}(1990)}]{Kip90}
{Kippenhahn}, R. \& {Weigert}, A. 1990, {Stellar Structure and Evolution}

\bibitem[{{Kiseleva} {et~al.}(1998){Kiseleva}, {Eggleton}, \&
  {Mikkola}}]{Kis98}
{Kiseleva}, L.~G., {Eggleton}, P.~P., \& {Mikkola}, S. 1998, \mnras, 300, 292

\bibitem[{{Kiseleva} {et~al.}(1994){Kiseleva}, {Eggleton}, \& {Orlov}}]{Kis94}
{Kiseleva}, L.~G., {Eggleton}, P.~P., \& {Orlov}, V.~V. 1994, \mnras, 270, 936

\bibitem[{{Kochanek}(1992)}]{Koc92}
{Kochanek}, C.~S. 1992, \apj, 385, 604

\bibitem[{{Kouwenhoven} {et~al.}(2007){Kouwenhoven}, {Brown}, {Portegies
  Zwart}, \& {Kaper}}]{Kou07}
{Kouwenhoven}, M.~B.~N., {Brown}, A.~G.~A., {Portegies Zwart}, S.~F., \&
  {Kaper}, L. 2007, \aap, 474, 77

\bibitem[{{Kozai}(1962)}]{Koz62}
{Kozai}, Y. 1962, \aj, 67, 591

\bibitem[{{Kroupa} {et~al.}(1993){Kroupa}, {Tout}, \& {Gilmore}}]{Kro93}
{Kroupa}, P., {Tout}, C.~A., \& {Gilmore}, G. 1993, \mnras, 262, 545

\bibitem[{{Lajoie} \& {Sills}(2011)}]{Laj11}
{Lajoie}, C.-P. \& {Sills}, A. 2011, \apj, 726, 67

\bibitem[{{Layton} {et~al.}(1998){Layton}, {Blondin}, {Owen}, \&
  {Stevens}}]{Lay98}
{Layton}, J.~T., {Blondin}, J.~M., {Owen}, M.~P., \& {Stevens}, I.~R. 1998,
  \na, 3, 111

\bibitem[{{Li} {et~al.}(2014){Li}, {Naoz}, {Holman}, \& {Loeb}}]{Li14}
{Li}, G., {Naoz}, S., {Holman}, M., \& {Loeb}, A. 2014, \apj, 791, 86

\bibitem[{{Lidov}(1962)}]{Lid62}
{Lidov}, M.~L. 1962, \planss, 9, 719

\bibitem[{{Lithwick} \& {Naoz}(2011)}]{Lit11}
{Lithwick}, Y. \& {Naoz}, S. 2011, \apj, 742, 94

\bibitem[{{Liu} {et~al.}(2015){Liu}, {Mu{\~n}oz}, \& {Lai}}]{Liu15}
{Liu}, B., {Mu{\~n}oz}, D.~J., \& {Lai}, D. 2015, \mnras, 447, 747

\bibitem[{{Lu} \& {Naoz}(2019)}]{Lu19}
{Lu}, C.~X. \& {Naoz}, S. 2019, \mnras, 484, 1506

\bibitem[{{Luo} {et~al.}(2016){Luo}, {Katz}, \& {Dong}}]{Luo16}
{Luo}, L., {Katz}, B., \& {Dong}, S. 2016, \mnras, 458, 3060

\bibitem[{{Maeder} \& {Meynet}(2000)}]{Mae00}
{Maeder}, A. \& {Meynet}, G. 2000, \araa, 38, 143

\bibitem[{{Malogolovets} {et~al.}(2007){Malogolovets}, {Balega}, \&
  {Rastegaev}}]{Mal07}
{Malogolovets}, E.~V., {Balega}, Y.~Y., \& {Rastegaev}, D.~A. 2007,
  Astrophysical Bulletin, 62, 111

\bibitem[{{Mardling} \& {Aarseth}(1999)}]{Mar99}
{Mardling}, R. \& {Aarseth}, S. 1999, in NATO Advanced Science Institutes (ASI)
  Series C, Vol. 522, NATO Advanced Science Institutes (ASI) Series C, ed.
  B.~A. {Steves} \& A.~E. {Roy}, 385

\bibitem[{{Mardling}(1995)}]{Mar95}
{Mardling}, R.~A. 1995, \apj, 450, 732

\bibitem[{{Mardling} \& {Aarseth}(2001)}]{Mar01}
{Mardling}, R.~A. \& {Aarseth}, S.~J. 2001, \mnras, 321, 398

\bibitem[{{Mazeh} \& {Shaham}(1979)}]{Maz79}
{Mazeh}, T. \& {Shaham}, J. 1979, \aap, 77, 145

\bibitem[{{McMillan}(1986)}]{McM86}
{McMillan}, S.~L.~W. 1986, \apj, 306, 552

\bibitem[{{Mennekens} {et~al.}(2010){Mennekens}, {Vanbeveren}, {De Greve}, \&
  {De Donder}}]{Men10}
{Mennekens}, N., {Vanbeveren}, D., {De Greve}, J.~P., \& {De Donder}, E. 2010,
  \aap, 515, A89

\bibitem[{{Michaely} \& {Perets}(2014)}]{Mic14}
{Michaely}, E. \& {Perets}, H.~B. 2014, \apj, 794, 122

\bibitem[{{Moe} \& {Di Stefano}(2017)}]{Moe17}
{Moe}, M. \& {Di Stefano}, R. 2017, \apjs, 230, 15

\bibitem[{{Moe} \& {Kratter}(2018)}]{Moe18}
{Moe}, M. \& {Kratter}, K.~M. 2018, \apj, 854, 44

\bibitem[{{Mohamed} \& {Podsiadlowski}(2007)}]{Moh07}
{Mohamed}, S. \& {Podsiadlowski}, P. 2007, Astronomical Society of the Pacific
  Conference Series, Vol. 372, {Wind Roche-Lobe Overflow: a New Mass-Transfer
  Mode for Wide Binaries}, ed. R.~{Napiwotzki} \& M.~R. {Burleigh}, 397

\bibitem[{{Naoz}(2016)}]{Nao16}
{Naoz}, S. 2016, \araa, 54, 441

\bibitem[{{Naoz} \& {Fabrycky}(2014)}]{Nao14}
{Naoz}, S. \& {Fabrycky}, D.~C. 2014, \apj, 793, 137

\bibitem[{{Naoz} {et~al.}(2013){Naoz}, {Farr}, {Lithwick}, {Rasio}, \&
  {Teyssandier}}]{Nao13}
{Naoz}, S., {Farr}, W.~M., {Lithwick}, Y., {Rasio}, F.~A., \& {Teyssandier}, J.
  2013, \mnras, 431, 2155

\bibitem[{{Nelemans} {et~al.}(2001){Nelemans}, {Yungelson}, {Portegies Zwart},
  \& {Verbunt}}]{Nel01}
{Nelemans}, G., {Yungelson}, L.~R., {Portegies Zwart}, S.~F., \& {Verbunt}, F.
  2001, \aap, 365, 491

\bibitem[{{Nicholls} \& {Wood}(2012)}]{Nic12}
{Nicholls}, C.~P. \& {Wood}, P.~R. 2012, \mnras, 421, 2616

\bibitem[{{Perets} \& {Kratter}(2012)}]{Per12}
{Perets}, H.~B. \& {Kratter}, K.~M. 2012, \apj, 760, 99

\bibitem[{{Perpiny{\`a}-Vall{\`e}s} {et~al.}(2019){Perpiny{\`a}-Vall{\`e}s},
  {Rebassa-Mansergas}, {G{\"a}nsicke}, {Toonen}, {Hermes}, {Gentile Fusillo},
  \& {Tremblay}}]{Per19}
{Perpiny{\`a}-Vall{\`e}s}, M., {Rebassa-Mansergas}, A., {G{\"a}nsicke}, B.~T.,
  {et~al.} 2019, \mnras, 483, 901

\bibitem[{{Peters}(1964)}]{Pet64}
{Peters}, P.~C. 1964, Physical Review, 136, 1224

\bibitem[{{Petrova} \& {Orlov}(1999)}]{Pet99}
{Petrova}, A.~V. \& {Orlov}, V.~V. 1999, \aj, 117, 587

\bibitem[{{Pijloo} {et~al.}(2012){Pijloo}, {Caputo}, \& {Portegies
  Zwart}}]{Pij12}
{Pijloo}, J.~T., {Caputo}, D.~P., \& {Portegies Zwart}, S.~F. 2012, \mnras,
  424, 2914

\bibitem[{{Portegies Zwart} \& {Leigh}(2019)}]{Por19}
{Portegies Zwart}, S. \& {Leigh}, N. W.~C. 2019, \apjl, 876, L33

\bibitem[{{Portegies Zwart} \& {McMillan}(2018)}]{Por18}
{Portegies Zwart}, S. \& {McMillan}, S. 2018, {Astrophysical Recipes; The art
  of AMUSE}

\bibitem[{{Portegies Zwart} {et~al.}(2009){Portegies Zwart}, {McMillan},
  {Harfst}, {Groen}, {Fujii}, {Nuall{\'a}in}, {Glebbeek}, {Heggie}, {Lombardi},
  {Hut}, {Angelou}, {Banerjee}, {Belkus}, {Fragos}, {Fregeau}, {Gaburov},
  {Izzard}, {Juri{\'c}}, {Justham}, {Sottoriva}, {Teuben}, {van Bever},
  {Yaron}, \& {Zemp}}]{Por09}
{Portegies Zwart}, S., {McMillan}, S., {Harfst}, S., {et~al.} 2009, \na, 14,
  369

\bibitem[{{Portegies Zwart} {et~al.}(2013){Portegies Zwart}, {McMillan}, {van
  Elteren}, {Pelupessy}, \& {de Vries}}]{Por13}
{Portegies Zwart}, S., {McMillan}, S.~L.~W., {van Elteren}, E., {Pelupessy},
  I., \& {de Vries}, N. 2013, Computer Physics Communications, 183, 456

\bibitem[{{Portegies Zwart} \& {van den Heuvel}(2016)}]{Por16}
{Portegies Zwart}, S.~F. \& {van den Heuvel}, E.~P.~J. 2016, \mnras, 456, 3401

\bibitem[{{Portegies Zwart} \& {Verbunt}(1996)}]{Por96}
{Portegies Zwart}, S.~F. \& {Verbunt}, F. 1996, \aap, 309, 179

\bibitem[{{Portegies Zwart} \& {Yungelson}(1998)}]{Por98}
{Portegies Zwart}, S.~F. \& {Yungelson}, L.~R. 1998, \aap, 332, 173

\bibitem[{{Press} \& {Teukolsky}(1977)}]{Pre77}
{Press}, W.~H. \& {Teukolsky}, S.~A. 1977, \apj, 213, 183

\bibitem[{{Raghavan} {et~al.}(2010){Raghavan}, {McAlister}, {Henry}, {Latham},
  {Marcy}, {Mason}, {Gies}, {White}, \& {ten Brummelaar}}]{Rag10}
{Raghavan}, D., {McAlister}, H.~A., {Henry}, T.~J., {et~al.} 2010, \apjs, 190,
  1

\bibitem[{{Raguzova} \& {Popov}(2005)}]{Rag05}
{Raguzova}, N.~V. \& {Popov}, S.~B. 2005, Astronomical and Astrophysical
  Transactions, 24, 151

\bibitem[{{Reg{\"o}s} {et~al.}(2005){Reg{\"o}s}, {Bailey}, \&
  {Mardling}}]{Reg05}
{Reg{\"o}s}, E., {Bailey}, V.~C., \& {Mardling}, R. 2005, \mnras, 358, 544

\bibitem[{{Remage Evans}(2011)}]{Rem11}
{Remage Evans}, N. 2011, Bulletin de la Societe Royale des Sciences de Liege,
  80, 663

\bibitem[{{Ruiter} {et~al.}(2009){Ruiter}, {Belczynski}, \& {Fryer}}]{Rui09}
{Ruiter}, A.~J., {Belczynski}, K., \& {Fryer}, C. 2009, \apj, 699, 2026

\bibitem[{{Saladino} {et~al.}(2018){Saladino}, {Pols}, {van der Helm},
  {Pelupessy}, \& {Portegies Zwart}}]{Sal18}
{Saladino}, M.~I., {Pols}, O.~R., {van der Helm}, E., {Pelupessy}, I., \&
  {Portegies Zwart}, S. 2018, \aap, 618, A50

\bibitem[{{Sana} {et~al.}(2012){Sana}, {de Mink}, {de Koter}, {Langer},
  {Evans}, {Gieles}, {Gosset}, {Izzard}, {Le Bouquin}, \& {Schneider}}]{San12}
{Sana}, H., {de Mink}, S.~E., {de Koter}, A., {et~al.} 2012, Science, 337, 444

\bibitem[{{Sana} {et~al.}(2014){Sana}, {Le Bouquin}, {Lacour}, {Berger},
  {Duvert}, {Gauchet}, {Norris}, {Olofsson}, {Pickel}, {Zins}, {Absil}, {de
  Koter}, {Kratter}, {Schnurr}, \& {Zinnecker}}]{San14}
{Sana}, H., {Le Bouquin}, J.-B., {Lacour}, S., {et~al.} 2014, \apjs, 215, 15

\bibitem[{{Scholz} {et~al.}(2004){Scholz}, {Lodieu}, {Ibata}, {Bienaym{\'e}},
  {Irwin}, {McCaughrean}, \& {Schwope}}]{Sch04}
{Scholz}, R.-D., {Lodieu}, N., {Ibata}, R., {et~al.} 2004, \mnras, 347, 685

\bibitem[{{Sepinsky} {et~al.}(2007{\natexlab{a}}){Sepinsky}, {Willems}, \&
  {Kalogera}}]{Sep07}
{Sepinsky}, J.~F., {Willems}, B., \& {Kalogera}, V. 2007{\natexlab{a}}, \apj,
  660, 1624

\bibitem[{{Sepinsky} {et~al.}(2007{\natexlab{b}}){Sepinsky}, {Willems},
  {Kalogera}, \& {Rasio}}]{Sep07b}
{Sepinsky}, J.~F., {Willems}, B., {Kalogera}, V., \& {Rasio}, F.~A.
  2007{\natexlab{b}}, \apj, 667, 1170

\bibitem[{{Sepinsky} {et~al.}(2009){Sepinsky}, {Willems}, {Kalogera}, \&
  {Rasio}}]{Sep09}
{Sepinsky}, J.~F., {Willems}, B., {Kalogera}, V., \& {Rasio}, F.~A. 2009, \apj,
  702, 1387

\bibitem[{{Seto}(2013)}]{Set13}
{Seto}, N. 2013, Physical Review Letters, 111, 061106

\bibitem[{{Shappee} \& {Thompson}(2013)}]{Sha13}
{Shappee}, B.~J. \& {Thompson}, T.~A. 2013, \apj, 766, 64

\bibitem[{{Smeyers} \& {Willems}(2001)}]{Sme01}
{Smeyers}, P. \& {Willems}, B. 2001, \aap, 373, 173

\bibitem[{{Soker}(2000)}]{Sok00}
{Soker}, N. 2000, \aap, 357, 557

\bibitem[{{Teyssandier} {et~al.}(2013){Teyssandier}, {Naoz}, {Lizarraga}, \&
  {Rasio}}]{Tey13}
{Teyssandier}, J., {Naoz}, S., {Lizarraga}, I., \& {Rasio}, F.~A. 2013, \apj,
  779, 166

\bibitem[{{Tokovinin}(2008)}]{Tok08}
{Tokovinin}, A. 2008, \mnras, 389, 925

\bibitem[{{Tokovinin}(2010)}]{Tok10}
{Tokovinin}, A. 2010, VizieR Online Data Catalog, 738

\bibitem[{{Tokovinin}(2014)}]{Tok14b}
{Tokovinin}, A. 2014, \aj, 147, 87

\bibitem[{{Tokovinin}(2016)}]{Tok16b}
{Tokovinin}, A. 2016, \apj, 831, 151

\bibitem[{{Tokovinin}(2017)}]{Tok17}
{Tokovinin}, A. 2017, \apj, 844, 103

\bibitem[{{Tokovinin} \& {Kiyaeva}(2016)}]{Tok16}
{Tokovinin}, A. \& {Kiyaeva}, O. 2016, \mnras, 456, 2070

\bibitem[{{Tokovinin} \& {Latham}(2017)}]{Tok17b}
{Tokovinin}, A. \& {Latham}, D.~W. 2017, \apj, 838, 54

\bibitem[{{Toonen} {et~al.}(2014){Toonen}, {Claeys}, {Mennekens}, \&
  {Ruiter}}]{Too14}
{Toonen}, S., {Claeys}, J.~S.~W., {Mennekens}, N., \& {Ruiter}, A.~J. 2014,
  \aap, 562, A14

\bibitem[{{Toonen} {et~al.}(2016){Toonen}, {Hamers}, \& {Portegies
  Zwart}}]{Too16}
{Toonen}, S., {Hamers}, A., \& {Portegies Zwart}, S. 2016, Computational
  Astrophysics and Cosmology, 3, 6

\bibitem[{{Toonen} {et~al.}(2017){Toonen}, {Hollands}, {G{\"a}nsicke}, \&
  {Boekholt}}]{Too17}
{Toonen}, S., {Hollands}, M., {G{\"a}nsicke}, B.~T., \& {Boekholt}, T. 2017,
  \aap, 602, A16

\bibitem[{{Toonen} {et~al.}(2012){Toonen}, {Nelemans}, \& {Portegies
  Zwart}}]{Too12}
{Toonen}, S., {Nelemans}, G., \& {Portegies Zwart}, S. 2012, \aap, 546, A70

\bibitem[{{Toonen} {et~al.}(2018{\natexlab{a}}){Toonen}, {Perets}, \&
  {Hamers}}]{Too18b}
{Toonen}, S., {Perets}, H.~B., \& {Hamers}, A.~S. 2018{\natexlab{a}}, \aap,
  610, A22

\bibitem[{{Toonen} {et~al.}(2018{\natexlab{b}}){Toonen}, {Perets}, {Igoshev},
  {Michaely}, \& {Zenati}}]{Too18}
{Toonen}, S., {Perets}, H.~B., {Igoshev}, A.~P., {Michaely}, E., \& {Zenati},
  Y. 2018{\natexlab{b}}, \aap, 619, A53

\bibitem[{{van den Berk} {et~al.}(2007){van den Berk}, {Portegies Zwart}, \&
  {McMillan}}]{Van07}
{van den Berk}, J., {Portegies Zwart}, S.~F., \& {McMillan}, S.~L.~W. 2007,
  \mnras, 379, 111

\bibitem[{{van der Helm} {et~al.}(2016){van der Helm}, {Portegies Zwart}, \&
  {Pols}}]{Hel16}
{van der Helm}, E., {Portegies Zwart}, S., \& {Pols}, O. 2016, \mnras, 455, 462

\bibitem[{{van Haaften} {et~al.}(2013){van Haaften}, {Nelemans}, {Voss},
  {Toonen}, {Portegies Zwart}, {Yungelson}, \& {van der Sluys}}]{Haa13}
{van Haaften}, L.~M., {Nelemans}, G., {Voss}, R., {et~al.} 2013, \aap, 552, A69

\bibitem[{{Van Winckel} {et~al.}(1995){Van Winckel}, {Waelkens}, \&
  {Waters}}]{Win95}
{Van Winckel}, H., {Waelkens}, C., \& {Waters}, L.~B.~F.~M. 1995, \aap, 293

\bibitem[{{Vos} {et~al.}(2015){Vos}, {{\O}stensen}, {Marchant}, \& {Van
  Winckel}}]{Vos15}
{Vos}, J., {{\O}stensen}, R.~H., {Marchant}, P., \& {Van Winckel}, H. 2015,
  \aap, 579, A49

\bibitem[{{Walter} {et~al.}(2015){Walter}, {Lutovinov}, {Bozzo}, \&
  {Tsygankov}}]{Wal15}
{Walter}, R., {Lutovinov}, A.~A., {Bozzo}, E., \& {Tsygankov}, S.~S. 2015,
  \aapr, 23, 2

\bibitem[{{Wegner} \& {McMahan}(1988)}]{Weg88}
{Wegner}, G. \& {McMahan}, R.~K. 1988, \aj, 96, 1933

\bibitem[{{Xu} {et~al.}(2015){Xu}, {Xia}, \& {Fu}}]{Xu15}
{Xu}, X.-B., {Xia}, F., \& {Fu}, Y.-N. 2015, Research in Astronomy and
  Astrophysics, 15, 1857

\bibitem[{{Yungelson} {et~al.}(1993){Yungelson}, {Tutukov}, \& {Livio}}]{Yun93}
{Yungelson}, L.~R., {Tutukov}, A.~V., \& {Livio}, M. 1993, \apj, 418, 794

\bibitem[{{Zavala} {et~al.}(2010){Zavala}, {Hummel}, {Boboltz}, {Ojha},
  {Shaffer}, {Tycner}, {Richards}, \& {Hutter}}]{Zav10}
{Zavala}, R.~T., {Hummel}, C.~A., {Boboltz}, D.~A., {et~al.} 2010, \apjl, 715,
  L44

\bibitem[{{Zavala} {et~al.}(2017){Zavala}, {Hummel}, {Boboltz}, {Ojha},
  {Shaffer}, {Tycner}, {Richards}, \& {Hutter}}]{Zav17}
{Zavala}, R.~T., {Hummel}, C.~A., {Boboltz}, D.~A., {et~al.} 2017, \apjl, 843,
  L18

\end{thebibliography}

\end{document}